\documentclass[a4paper,english,latin9]{aa}
\usepackage{lmodern}

\usepackage[T1]{fontenc}
\usepackage{float}
\usepackage{textcomp}
\usepackage{amsmath}
\usepackage{graphicx}
\usepackage[varg]{txfonts}

\makeatletter

\pdfpageheight\paperheight
\pdfpagewidth\paperwidth

\providecommand{\tabularnewline}{\\}

\@ifundefined{showcaptionsetup}{}{%
 \PassOptionsToPackage{caption=false}{subfig}}
\usepackage{subfig}
\makeatother

\usepackage{babel}
\begin{document}
\title{}
\title{Assessing the potential for liquid solvents from X-ray sources:\\
 considerations on bodies orbiting AGN }
 \titlerunning{The potential for liquid solvents from X-ray sources}
\author{D. Rodener\inst{1}, M. Hausmann\inst{1}\and G. Hildenbrand\inst{1}}
\offprints{G. Hildenbrand}
\institute{$^{1}$Kirchhoff-Institute for Physics, Heidelberg University, INF
227, 69117 Heidelberg, Germany.}
\date{Received XXX ; accepted XXX}
\abstract{}{We aim to establish a rough first prospect on the potential of certain
biorelevant solvents (water, ammonia and methane) to be present in
liquid form inside the uppermost few meters of several modelled surfaces
(rocky and icy crusts of various compositions) of hypothetical bodies
orbiting active galactic nuclei (AGN), and investigate under which
constraints this might occur.}{We adjust and average together X-ray spectra from a sample of 20 Type
1 Seyfert galaxies to calculate a mean snowline of the sample used.
Based on this, we introduce variation of a hypothetical body's orbit
across distances between 10\% and 100\% of the snowline radius, and
calculate a sub-surface attenuation within four different model surface
compositions for each. Surface compositions are based on lunar soil
and solvent ices found in the milky way's circumnuclear region. We
then use this as a continuous source term for a thermal model. Example
bodies are systematically investigated with sizes between $1/30$
and $20$ earth radii. Further outlier variations are also considered
(such as the case of bound rotation of the body) to end up with a
perspective of solvent phases under a wide slew of many different
conditions.}{We find that liquid solvents are possible under a multitude of parameters,
with temperature being the main constraint to liquid water and body
size as well as pressure being the main constraint to liquid methane
and ammonia. We further find that these results, when adjusted for
snowline distance, depend less on the energy output of the central
source within the Seyfert Type 1 AGN than on other parameters, such as body
sizes and solvent properties.}{}
\keywords{galaxies: active \textendash{} galaxies: Seyfert \textendash{} galaxies:
nuclei \textendash{} X-rays: galaxies \textendash{} planets and satellites:
surfaces \textendash{} astrobiology}
\maketitle

\section{Introduction}

We aim to investigate some of the factors that play into the potential
of (and the process of detecting) life outside of earth's own biosphere.
We do this using a rudimentary first look into constraints of the
habitability regarding one of the most bizarre astrophysical environments
imaginable: the circumnuclear regions of active galactic nuclei.

The pre-existing variety of research pertaining to energy sources
and the formation of liquid environments was an important step for
astrophysics, astrochemistry as well as astrobiology. However, the
respective research of uncommon energy sources has so far not thoroughly
and in depth touched upon the possibilities that the combination of
sub-surface environments and high-energy radiation offer.

Considerations on habitability get exceedingly complex, encompassing
a wealth of topics such as energy supply, mechanisms to gather, store,
transport, and process said energy in biochemical form, or the way
by which extraterrestrial organisms might shield themselves from harmful
influences such as radiation or temperature extremes. As such, a comprehensive
look at the habitability of the environments discussed here would
go far beyond any resonable scope. We focus on only one aspect of
habitability that both seems to be the most prevalent in both standard
and proposed exotic biologies and is expected to exist in great abundance
in the universe: tiquid solvents.

The solvents were chosen based on the following considerations: water
goes almost without saying, as it is the key solvent for most life
on earth and fulfills many important roles such as providing a transportation
medium to distribute molecules throughout the organism. Methane and
ammonia are also important solvents for complex life on earth, and
are furthermore proposed in some models for exotic biochemistries
(\citealt{Stevenson2015}, \citealt{Palmer2017}, \citealt{Rampelotto2010})
with liquid methane proven to exist on Saturn's moon titan (\citealt{Stofan2007})
and liquid ammonia being suspected to (\citealt{Grasset2000}). These
promising findings, as well as the beneficial thermochemical impact
of salt-mixtures on water (with $NaCl$ being the most common salt
in water on earth and $CaCl_{2}$ showing the most beneficial impact),
lead us to choose these five solvents for our investigation.

\begin{figure*}
\begin{centering}
\includegraphics[width=0.95\textwidth]{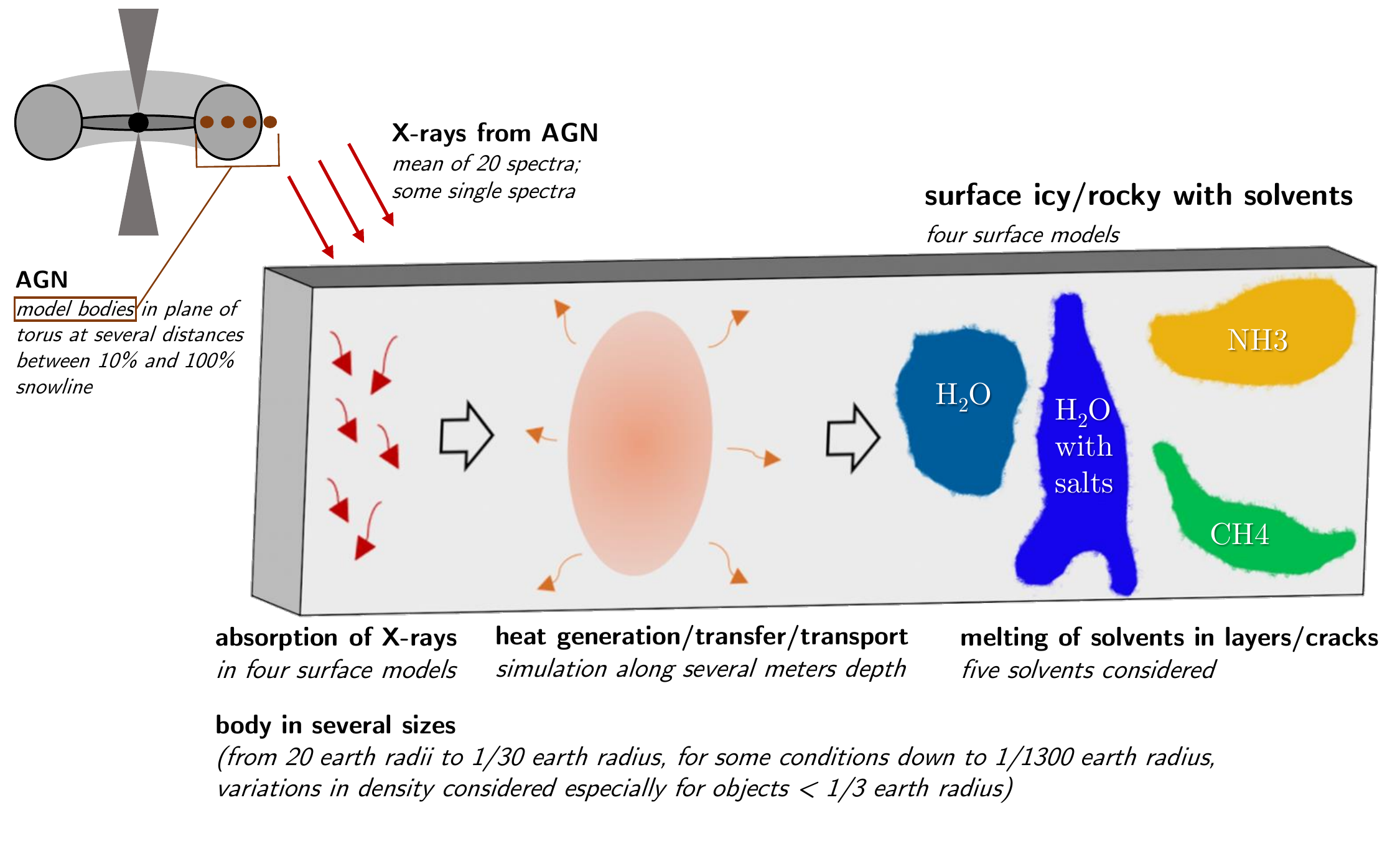}
\par\end{centering}
\caption{\label{fig:Sketch}Outline of the process behind this work: The mean
of X-ray energy spectra of 20 individual AGN is used to compute energy
absorption within five surface models.}
\end{figure*}

Regarding the AGN, this work will thusly focus mainly on radio-quiet
ones as external energy sources with X-ray emissions strong enough
for their radiation to mostly penetrate both otherwise optically thick
circumnuclear material\footnote{similar to the reasoning behind using X-ray luminosity for snowline
calculations (\citealt{Wada2019,Wada2021})} and the upper regions of regolith on bodies in their vicinity (which
encompasses several parsec for objects such as these). This coincides
with a recent proposal of planetary formation in the snowline around
Seyfert-type AGN (\citealt{Wada2019}), but despite this overlap,
we want to avoid strict adherence to the planetary properties considered
in that work to keep a broad perspective on this uncommon field and
check a reasonably wide slew of possibilities. A graphic overview
of the model is given in Fig. \ref{fig:Sketch}.

\section{Methods}

\subsection{Obtaining a mean energy spectrum}

We sample 20 Type 1 Seyfert galaxies taken from a multi-wavelength
catalogue of such objects. (\citealt{AGNSEDATLAS})

Erroneous short dips into negative flux values observed in the data
were interpreted as errors, likely of the detector, and were replaced
with the value of the longer wavelength point adjacent to the sudden
dip. Given the so-replaced values lie well within the bulk of the
dataset, and that this was only necessary for a single spectrum, this
solution was deemed satisfactory.

To obtain appropriate liquid layers in the subsurface, X-rays need
to penetrate a non-negligble depth. We therefore focus on X-ray photon
energies in the dataset between $1$ and $120keV$, as photons below
$1keV$ do not penetrate very deep and the respective integrated flux
does not generate enough heat to be of significance for the overall
result of deeper liquid subsurface layers.

After having obtained such a spectra, a sketch of the modelling process
is given in Fig. \ref{fig:Sketch}: The calculated absorbed energy
is used as continious source term for a thermal model calculating
heat generation, transfer and transport down to a depth of 3 m, at
the end of which a temperature profile is generated. This temperature
profile is checked against the thermochemical properties of five model
solvents (in combination with a value of pressure at different depths,
calculated from 9 different body configurations) to determine where
a given solvent could exist in the liquid phase.

\subsubsection{Adjusting measurements for distance and angle}

Using the distances (their calculation is outlined in Appendix \ref{sec:Obtaining-distances-to})
for the observed targets from the point of observation (earth) $R_{earth}$,
which were obtained using the astroquery python package (\citealt{Astroquery}),
as well as the flux received at this point of observation $S_{earth}$,
we can determine the flux $S_{model}$ received at a model planet
at an orbit of radius $r_{model}$ using the inverse square law to:

\noindent
\begin{equation}
S_{model}=\frac{S_{earth}}{(\frac{r_{model}}{R_{earth}})^{2}},\label{eq:DistAdjust}
\end{equation}
However, this is first done for a distance of $10pc$ from the central
source to establish a point at which we can calculate the mean of
the taken sample.

Viewing angle is a more delicate issue. We can expect that most AGN
show an incredibly large energy emission in pole-on direction (viewing
the accretion disk from above) and only marginal energy emission in
disk-on direction (\citealt{Padovani2017}), however, we are unable
to fully verify this as we are unable to observe a single such object
from multiple angles. This is further complicated by a lack of precise
information about the viewing angle at which we see these objects,
as AGN are both too bright and too far away to resolve them sufficiently
to make geometric assumptions about viewing angle, instead having
to resort to using kinematics (\citealt{Fischer_2013}) or similar
derivation techniques. To combat these problems here, we use the assumption
of a simple unified model of AGNs, which poses that different types
of active galactic nuclei are all similar objects viewed at different
angles and at different stages of a similar evolutionary process.(\citealt{Antonucci1993})
According to this assumption then, Seyfert 1 galaxies are viewed as
pole-on, derived from the lack of broad emission lines that would
be imbued in the spectrum by region of high-velocity dust and gas
surrounding the accretion disk close to the equatorial plane.

However, a planet forming in the circumnuclear disk would not exist
this high above the disk\textquoteright s equatorial plane, it would
be much more likely to form at a lower angle close to the denser regions
of dust surrounding the AGN. To adjust for this in a sufficiently
concise way, we utilize a formula expressing the reduction of observed
apparent luminosity for a unified luminosity function for AGNs from
(\citealt{Zhang2004}), which expresses a reduction of the observed
apparent luminosity by a factor of:

\noindent
\begin{equation}
A_{\theta}=\cos\theta*(\frac{1+2\cos\theta}{3}),\label{eq:AngleAdjust}
\end{equation}
which then leads to a formula for the observed luminosity $I'_{\lambda}$:

\noindent
\begin{equation}
I'_{\lambda}=I_{\lambda}*A_{\theta},\label{eq:AngleAdjust-1}
\end{equation}

\noindent where $\theta$ is the inclination angle of the accreation
disk relative to the observer, with $\theta=90{^\circ}$ for an edge-on
view of the disk.

As the same work classifies Type 2 AGNs at an angle of 68\textdegree{}
or greater, this angle would correspond to a position on the edge
of the broad line region of the circumnuclear disk, and will therefore
be used as an example value to set the model in.

\subsubsection{Interpolating and Averaging}

To effectively calculate a mean of the Seyfert galaxy sample, which
all exhibit unevent datapoints, a universal grid is defined for the
energy axis of the spectra that runs from the first to the last photon
energy value that is in at least one of the spectral grids used. In
this case this is between $1$ and $120keV$, defined with 5205 logarithmically
equidistant steps. (The exact number arose from the construction of
the framework.) This universal grid is then sliced corresponding to
each of the contributing spectra, and each spectrum is in turn interpolated
onto its respective slice. Interpolation is done using the ``interp1d''-function
of python's ``numpy''-package (\citealt{2020NumPy-Array}), set
to also extrapolate missing values to circumvent conflicts at the
boundaries. Values outside these boundaries (above and below a certain
spectrum's recorded range) are set to NaN. The result of this is an
array of 20 spectra along a single x-axis.

After transposing, the mean can be calculated per energy value (in
a kind of cross-section of the spectra) using numpy's ``nanmean''-function,
which automatically disregards any NaN-values. This outputs a mean
Seyfert 1 spectrum as it would be observed at a distance of 10 pc.

The distance at which this mean spectrum would exhibit a snowline
is defined as the distance from the source where the equilibrium temperature
of a body:

\noindent
\begin{equation}
T_{eq}=(\frac{I_{0}(1-A_{B})}{4\sigma})^{\frac{1}{4}},\label{eq:TempEqui}
\end{equation}

where $I$ denotes the energy flux, $A_{B}$the surface albedo and
$\sigma$ is the Stefan-Boltzmann constant as obtained using python's
scipy package (\citealt{2020SciPy-NMeth}), becomes a set temperature,
conventionally $170K$. In this case, the X-ray albedo is almost zero
and can be neglected, and we are interested at the flux at a certain
distance $I(r_{snow})=I_{0}$.

\begin{figure*}
\begin{centering}
\includegraphics[width=1\textwidth]{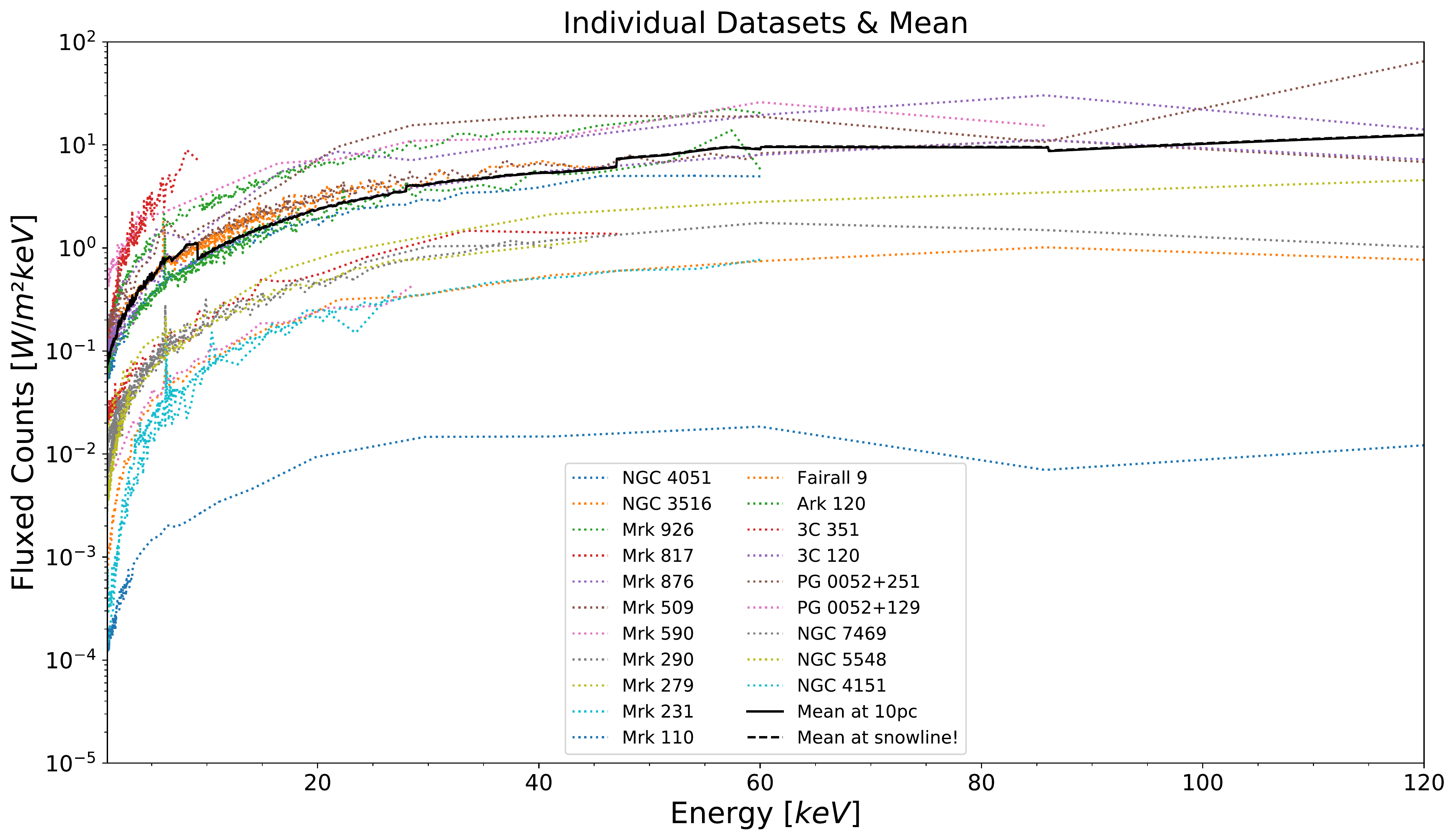}
\par\end{centering}
\caption{\label{fig:Spectra}Data of individual X-ray spectra based on the
catalogue of (\citealt{AGNSEDATLAS}) (between 1 and 120 keV) of 20
Seyfert 1 AGNs used in this work (colored and dotted) as they would
be observed at a distance of $10pc$, as well as a mean their spectra
as it would be observed a distance of $10pc$ (black, solid line)
and at a snowline distance of $11.45pc$ (black, dashed line).}
\end{figure*}

This snowline is calculated from the integrated flux of the mean spectrum
(done using numpy's trapezoid integration function integrate.trapz)
to be at a distance of $11.45pc$. Individual spectra as well as the
averaged spectrum at $10pc$ and the latter also at the snowline of
$11.45pc$ can be seen in Fig. \ref{fig:Spectra}.

Distance variation is introduced here with all following computations
being done for a set of ten fractions (from 10\% to 100\%) of the
snowline distance.

\subsection{Obtaining surface properties for model bodies}

With the radiation environment figured out, we construct models for
potential bodies as well as their uppermost few meters of crust. We
choose a series of body sizes between $20$ and $\frac{1}{30}R_{earth}$\footnote{For the bulk of the simulations. Special calculations with systems
as small as $\frac{1}{1300}R_{earth}$were done as well.}to systematically cover an array of possibilities, with two main achor
points: the factor $20$ (although originally attributed to earth
masses, not radii) stems from the planetary models proposed in (\citealt{Wada2021}),
and radii down to $\frac{1}{30}R_{earth}$ act as an attempt in reaching
sizes at which accretion energy and radioactive heating become less
important, equivalent to asteroids or tiny moons in the solar system.

\subsubsection{Proposing surface models\label{subsec:Proposing-surface-models}}

We lack chemical data about the circumnuclear disks (CNDs) of any
Seyfert-type galaxy, and information about exoplanetary surfaces are
extremely scarce as well.

To compensate for this, and build a reasonable initial model, we work
with chemical composition and morphology similar to lunar soil as
reported by (\citealt{Alexander2016}) to emulate rocky bodies and
a pure solvent ice regolith to emulate icy bodies without many minerals
in the upper layers, similar to icy moons found in the solar system.
The former was based on multiple points of reasoning: the moon is
the best-researched body without a significant atmosphere, we have
extensive geological data about its surface regolith and the few exoplanetary
surfaces we have data about have shown to be similar to the moon\textquoteright s
composition. (\citealt{Kreidberg2019})

To introduce solvents to the originally dry lunar model, we propose
the regolith porosity of 50\% to be filled with (initially) solvent
ices of water, methane, and ammonia. The ratio of these solvents,
both when introduced to the lunar model and as part of the icy model,
was determined \textendash{} as a loose guideline \textendash{} by
the chemical abundances obtained from spectroscopic observations and
models of the milky way\textquoteright s own central region (\citealt{Harada2015}).
Saltwater concentrations were chosen to yield maximum reduction of
melting points. End results of the rocky crust filled with freshwater
will be displayed in \ref{fig:RockyFreshLiquid}, while results for
the other models will be elaborated upon in Appendix \ref{sec:Exhaustive-model-simulations}.

In the interest of the scope of this work \textendash{} and because
minute details in the composition ultimately do not impact the mass
attenuation coefficient a lot \textendash{} we focus on the abundances
of the relevant solvents, ignoring other volatiles with generally
lower abundances such as $CO$ or $N_{2}$. The abundances of all
molecules involved in our four models are shown in Table \ref{tab:Abundances}.

\begin{table}
\caption{\label{tab:Abundances}Surface compositions}

\begin{centering}
\begin{tabular}{lllll}
\hline
Formula & \multicolumn{4}{c}{Abundances}\tabularnewline
\hline
 & Icy Composition & \multicolumn{3}{l}{Rocky Composition}\tabularnewline
\hline
 & Ice & Fresh$^{\textrm{a}}$ & $NaCl^{\textrm{b}}$ & $CaCl_{2}{}^{\textrm{c}}$\tabularnewline
\hline
$H_{2}O$ & 0.762 & 0.381 & 0.293 & 0.263\tabularnewline
$CH_{4}$ & 0.17 & 0.085 & 0.085 & 0.085\tabularnewline
$NH_{3}$ & 0.068 & 0.034 & 0.034 & 0.034\tabularnewline
\hline
$SiO_{2}$ &  & 0.25 & 0.25 & 0.25\tabularnewline
$FeO$ &  & 0.135 & 0.135 & 0.135\tabularnewline
$Al_{2}O_{3}$ &  & 0.06 & 0.06 & 0.06\tabularnewline
$CaO$ &  & 0.055 & 0.055 & 0.055\tabularnewline
\hline
$NaCl$ &  &  & 0.088 & \tabularnewline
$CaCl_{2}$ &  &  &  & 0.118\tabularnewline
\hline
\end{tabular}
\par\end{centering}
$^{\textrm{a}}$Lunar soil model with 50\% solvent ices \& fresh water

$^{\textrm{b}}$Lunar soil model with 50\% solvent ices \& 23 wt\%
$NaCl$-saltwater

$^{\textrm{c}}$Lunar soil model with 50\% solvent ices \& 31 wt\%
$CaCl_{2}$-saltwater
\end{table}

Another important factor (more important than even the input energy
as shown in Fig. \ref{fig:EnergAbs}) is surface density. We approximate
the overall complicated geometry of regolith by utilizing a calculated
density from the density of compounds involved\footnote{with data taken from the NIST WebBook (\citealt{NISTWebbook})}
(and the porosity assumed), which can be seen in Table \ref{tab:SurfDensity}.

\begin{table}
\caption{\label{tab:SurfDensity}Surface densities}

\centering{}%
\begin{tabular}{ll||l||l||l}
\hline
Surface Model & \multicolumn{4}{c}{Density $[\frac{g}{cm{{}^3}}]$}\tabularnewline
\hline
Rocky, freshwater & \multicolumn{4}{c}{$2.2765$}\tabularnewline
Rocky, $NaCl$-saltwater & \multicolumn{4}{c}{$2.3788$}\tabularnewline
Rocky, $CaCl_{2}$-saltwater & \multicolumn{4}{c}{$2.4121$}\tabularnewline
Icy & \multicolumn{4}{c}{$0.8929$}\tabularnewline
\hline
\end{tabular}
\end{table}

\subsubsection{Calculating attenuated energy using Lambert-Beer's law\label{subsec:Calculating-attenuated-energy}}

Attenuation in matter was calculated using Lambert-Beer's law in the
form:
\begin{equation}
I_{\lambda}=I_{0,\lambda}\exp(-\frac{\mu}{\rho}\rho d),
\end{equation}
where $I_{0,\lambda}$and $I_{\lambda}$ denote intensity respective
before and after traversing a depth $d$ inside matter of density
$\rho$ and with a mass attenuation coefficient of $\frac{\mu}{\rho}$.
For the photon energies considered, the mass attenuation coefficient
will vary with the photon energy. For this purpose, the mass attenuation
coefficient of the investigated surface compositions has been obtained
in an energy range between $1keV$ and $120keV$ using the national
institute of standards and technology (NIST) \citeauthor{NISTcalc}
database\textquoteright s online mass attenuation coefficient calculator
for mixtures of molecules, which has in turn interpolated to fit onto
the generalized energy grid constructed earlier. The result is shown
in Appendix \ref{sec:Mass-attenuation-coefficient}.

\begin{figure*}
\begin{centering}
\includegraphics[width=1\textwidth]{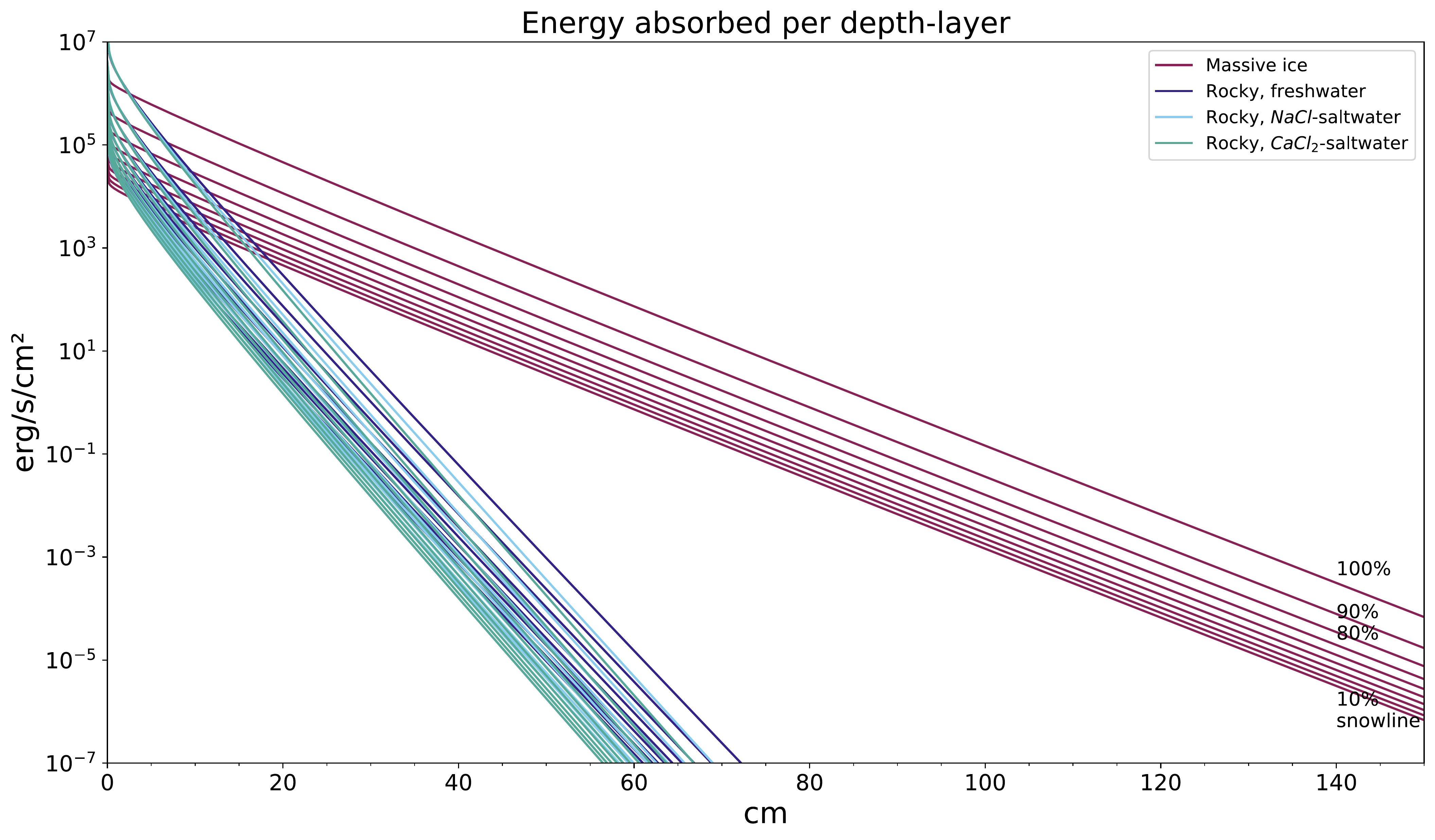}
\par\end{centering}
\caption{\label{fig:EnergAbs}Energy flux absorbed at a certain depth in 5
surface models, with a family of graphs from 10\% to 100\% snowline
distance (descriptions of which are only shown for pure ice models
for improved legibility in the graph), with different water-mixtures
being very close to each other in the rocky model. What can be clearly
seen is that salt water compositions have a small but noticeable impact
on absorption, although the main determining factor of the absorption
is material density. }
\end{figure*}

This creates a family of spectral attenuation curves over both the
energy and the depth grid. Integrating over the entire spectrum at
each depth-point $n$ allows us to obtain the bolometric intensity
per depth, which, when substracted from the intensity at a point $n-1$,
leads to information about the energy deposited inside the layer between
$n$ and $n-1$ and thusly what amount of energy is available for
heat generation\footnote{Only 90\% of which will be used to generate the heat in this model
however, with 10\% being ``reserved'' for effects not inspected
closer here such as secondary radiation and the modification of bonds
of chemical compounds. This is discussed in detail in Appendix \ref{sec:About-heat-conversion}.}. This is shown in Fig. \ref{fig:EnergAbs}.

\subsubsection{Preparing thermal properties: Triple point depth\label{subsec:Preparing-thermal-properties:}}

To determine whether a solvent is present in its liquid phase, the
two key factors are temperature and pressure. Both of which are a
complex matter which will need to be approximated to a certain degree,
starting with pressure here. We obtained these pressures $p_{triple}$
from the NIST Chemistry WebBook, NIST Standard Reference Database
Number 69 (\citealt{NISTWebbook}) and used them to calculate, for
each model body, the depth at which the triple point would be reached,
using:
\begin{equation}
d_{triple}=\frac{3p_{triple}}{4G\pi R_{body}\rho_{body}\rho_{surface}},\label{eq:TripleDepth}
\end{equation}

whereas $R_{body}$ and $\rho_{body}$denote the properties of the
model body, that being a radius between 1/30 and 20 times earth's
radius and a density equal to the overall average density of earth
($\rho_{body}=5.51\frac{g}{cm^{3}}$). $\rho_{surface}$describes
the density of the surface model as built in section \ref{subsec:Proposing-surface-models}.
A detailed derivation of (\ref{eq:TripleDepth}) is given in Appendix
\ref{subsec:Derivation-of-triple}.

\subsubsection{Vapor pressure curve}

The last step in determining the phase of a solvent, as mentioned
in the previous section, is temperature, namely where a solvent reaches
melting and boiling points. The melting point can be handled with
relative ease, as most solvents show a constant melting point equalling
the temperature at their triple point (anomalies notwithstanding).
Triple point temperatures are discussed in detail in Appendix \ref{sec:About-the-triple}.

To deal with the boiling point, we have access to multiple equations
approximating part of the phase profile of certain substances, with
a particularly simple and effective one being the Antoine equation:

\begin{equation}
\log_{10}p=A-\frac{B}{C+T},
\end{equation}

which relates the vapor pressure $p$ with the respective boiling
temperature $T$ and a set of empirically determined coefficients
$A,B\,\&\,C$. This can be easily rearranged to determine the temperature
instead of pressure with: $T=\frac{B}{A-\log_{10}p}-C$. Coefficients
for freshwater, ammonia and methane were obtained from the NIST Chemistry
WebBook, where we chose the sets with the widest temperature coverage,
and can be seen in Table \ref{tab:AntoineCoeffs}. The saltwaters
saw adjustments in the form of shifting the equation by about four
Kelvin along the T axis for both mixtures\footnote{corresponding to the boiling point shift of these mixtures at these
concentrations}.

\begin{table}
\caption{\label{tab:AntoineCoeffs}Antoine coefficients of utilized solvents}

\begin{centering}
\begin{tabular}{lccc}
\hline
Formula & \multicolumn{3}{c}{Coefficients$^{\textrm{a}}$}\tabularnewline
\hline
 & A & B & C\tabularnewline
\hline
$H_{2}O$ & $1435.264$ & $4.6543$ & $-64.848$\tabularnewline
$CH_{4}$ & $443.028$ & $3.9895$ & $-0.49$\tabularnewline
$NH_{3}$ & $506.713$ & $3.18757$ & $-80.78$\tabularnewline
\hline
\end{tabular}
\par\end{centering}
$^{\textrm{a}}$NIST Chemistry WebBook, NIST Standard Reference Database
Number 69 (\citealt{NISTWebbook})
\end{table}

\subsection{Simulating thermal profile within the surface}

\subsubsection{Calculating specific heat capacities using polynomial equations}

The goal is to arrive at a continuous temperature source term that
can be fed to a numerical thermal model. For this purpose, we can
compute temperature from the input energy derived earlier by using
the specific heat capacity
\begin{equation}
c_{p}=\frac{1}{m}*\frac{dQ}{dT},
\end{equation}
which allows us to obtain the temperature change $dT$ per timestep
of a mass element $m$ hit by a change in heat energy $dQ$, or in
this case an amount of radiative energy $I$ per timestep. We also
introduce the conservative approximation that only roughly 90\% of
radiation hitting an average substance will be converted into heat,
with the other 10\% leading to secondary radiation effects not further
discussed in this work. A detailed breakdown of the reasoning behind
this can be found in Appendix \ref{sec:About-heat-conversion}. Thus,
to calculate the source term $C$ we use:
\begin{equation}
C=dT=\frac{0.9*dQ}{m*c_{p}}=\frac{0.9*I}{V*\rho*c_{p}},
\end{equation}
splitting the mass element into volume $V$ and density $\rho$. Density
is carried over from the surface model constructed earlier, the radiation
flux from the adjusted, meaned and integrated photometric data. The
volume is chosen as a column of the height of one depth grid cell
($0.1cm$) over a unit area $1m^{2}$. This leaves the specific heat
capacity to be determined.

\emph{Specific heat capacity}: A popular way to account for temperature
dependency of the specific heat capacity is the usage of the \textquotedblleft Shomate
equation\textquotedblright{} , a polynomial equation that uses empirically
determined coefficients $A,B,C,D$ to approximate a $c_{p}$-$T$-curve.
We used a slightly modified polynomial equation taken from the Chemical
Engineering and Materials Research Information Center (\citeauthor{CHERIC})
database in the form of:

\begin{equation}
c_{p}=A+BT+CT^{2}+DT^{3},\label{eq:Shomate}
\end{equation}
with the result being in units of $\frac{kJ}{kg-molK}$. Noteworthy
is that kg-mol is a distinct unit equivalent to one kilomole, and
thusly $\frac{kJ}{kg-molK}=\frac{J}{molK}$. This library, however,
only encompasses values for liquids and gases, so solids need to be
treated with exceptions:

\emph{a) Rocky solids:} For most monoatomic solid materials of heavier
atoms, the specific heat capacity is relatively constant, following
the Dulong-Petit law (\citealt{LandauLifshitz1980}):

\begin{equation}
c_{p}=3R\approx24.9\frac{J}{molK},
\end{equation}

\emph{b) Ammonia ice \& methane ice:} For ammonia ice and methane
ice, tables containing experimental information about the specific
heat capacity can be found. For ammonia, this is Table IV in (\citealt{Overstreet1937})
which is used here up until 191 Kelvin, the melting point. For methane,
Table 2 from (\citealt{Yakub2016}) is used. In both cases, the so
obtained data is fitted onto (\ref{eq:Shomate}) using python\textquoteright s
SciPy package and a least square fit. The polynomial coefficients
obtained thusly are given over to the function calculating the overall
specific heat capacity.

\emph{c) Water ice: }For water ice, the specific heat is approximated
using a different formula specific to it, as obtained from (\citealt{Shulman2004})
in equation (1) of that work. It should be noted that this work outlines
more precise ways to approximate ice\textquoteright s specific heat
(that is the main purpose of said paper), but for the scope of this
work here, the lower precision formula given at the start is wholly
sufficient:
\begin{equation}
c_{p}\approx7.8*10^{-3}T\frac{J}{gK}*18\frac{g}{mol}=0.1404*T\frac{J}{molK}.
\end{equation}

With the specific heat capacities obtained, a function was written
to calculate the overall heat capacity of the mixture, depending on
current temperature and pressure of the simulation, using the known
abundances as a sum over all involved substances:

\begin{equation}
c_{p,tot}=\sum_{i}N_{i}*c_{p,i}.
\end{equation}

\subsubsection{Setting up a solver for the 1-D heat equation}

The core of this thermal solver is the 1-D heat equation with advection
and an additional continuous source term. A finite difference method
to solve this problem numerically was described in (\citeauthor{Riflet??})
based on a Forward Time Central Space (FTCS) scheme used to solve
the 1-D heat equation with no source term. This method utilized the
algorithm used to solve a 1-D heat equation with diffusivity and decay:

\noindent
\begin{equation}
T_{i}^{*n+1}=Dif*T_{i+1}^{n}+(1-k\Delta t-2Dif)T_{i}^{n}+Dif*T_{i-1}^{n}.
\end{equation}
Here, $T$ describes the temperature, $k$ is the decay coefficient
which will be set to 0\footnote{We do not account for lateral thermal decay as we are building a 1-D
model, and horizontal thermal decay only occurs at the surface, which
is controlled here using the equilibrium temperature.} and $\Delta t$ is the width of steps within the discrete time grid
used. In accordance with this, $\Delta x$ describes the spacing of
the space grid. Subscript indices (i) denote steps over the space
grid, while superscript indices (n) denote time. $Dif$ is the thermal
diffusivity coefficient adjusted onto the space and time grid used,
as:

\noindent
\begin{equation}
Dif=\mu\frac{\Delta t}{\Delta x^{2}},
\end{equation}
with thermal diffusivity coefficient $\mu$. Given the simplicity
of this scheme, we set $\mu=\mu_{ice}\approx1.02\frac{mm{{}^2}}{s}$.

To now implement continuous emission, a source term $c$ is added
flat to all spatial grid points $i=p$ where the external heating
applies:

\begin{eqnarray}
T_{i}^{n+1}= & T_{i}^{*n+1}\label{eq:Source1}\\
T_{i}^{n+1}= & T_{i}^{*n+1}+C\Delta t\label{eq:Source2}
\end{eqnarray}

In our case, the source term is applied to all spatial grid points
but varies over it. Within the code, this is expressed using Numpy
arrays along the spatial grid for both $T$ and $C$.

Furthermore, to implement a day-and-night cycle relatively easy, the
source term was only applied every other time step. Compared to a
simple one-half multiplier on the source term, this implementation
allows for the model to cool down during the simulated night steps.

\emph{Boundary conditions:} At the initial spatial grid point (the
exact surface layer), we set the temperature to be equal to the equilibrium
temperature given the radiation input at that point.

\noindent
\begin{equation}
T_{0}^{n+1}=T_{eq},
\end{equation}

The equilibrium temperature is an accurate measure for surface temperatures
of atmosphere-less bodies, hence why we deemed this boundary condition
a sound assumption. Otherwise, heat conducted upwards within the model
and reaching the surface would either build up indefinitely (due to
the lack of any further conduction and decay) or decay too fast (were
the decay term kept for this purpose).

At the initial time grid point (time 0) we set the entire temperature
profile to be the equilibrium temperature of the radiation influx
profile.

\noindent
\begin{equation}
T_{i}^{0}=T_{eq,i},
\end{equation}

This alone would not be an accurate thermal model, but it provides
a sufficient starting point. As the simulation progresses, conductive
effects will correct this initial outset quickly to a more realistic
profile.

\subsubsection{Forming a frame for the simulation with continuous input}

The full function takes arguments for the surface model, body size,
the grid spacing $\Delta x$ and $\Delta t$ as well as the full simulation
timeframe $time$. With a given profile of absorbed energy calculated
in section \ref{subsec:Calculating-attenuated-energy}, the boundary
conditions for the thermal solver are calculated and then applied
to that solver recursively, meaning each timestep$(T_{i}^{n})$ provides
the basis for the next timestep $(T_{i}^{n})$, over the length of
a while loop running until a counting variable $t$ reaches the full
time $time$. A day-and-night cycle is simulated by setting up a boolean
variable $day=True$ before initializing the loop, and then using
each loop step to flip the value using $day=not\:day$. The source
term is only applied when $day=True$, thusly only ever other step.

Results are given to a dictionary. Said dictionary is plotted directly\footnote{Calculations were done in seperate clusters, but plotting was done
in the general IPython environment. This necessitates extra steps
at this point that are laid out in more detail in section \ref{subsec:Notes-about-IPython}
.} and, using temperature and pressure criteria discussed in section
\ref{subsec:Preparing-thermal-properties:} and Appendix \ref{sec:About-the-triple},
used to evaluate the phase of contributing solvents.

\subsection{Notes about IPython multiprocessing for the simulation\label{subsec:Notes-about-IPython}}

The character of this work, with several simulations running over
multiple, independent parameters, lends itself to multiprocessing,
that is having multiple simulations run simultaneously on different
CPU threads. For this purpose, computations have been done in the
IPython environment (\citealt{IPython}) and use its built-in multiprocessing
package ipyparallel (\citealt{ipyparallel}). This allows us to execute
certain cells of IPython notebooks on different clusters, using a
multicore processor to its full extent while shortening overall computation
time if multiple simulations are run simultaneously.

With this arise some quirks: We ran all computations, with the exception
of the final simulation, on all clusters to ensure all necessary variables
were available in all clusters. The final plots were done in the general
IPython environment (using the matplotlib python package(\citealt{matplotlib})),
and for this purpose results from the simulation as well as necessary
variables from earlier parts of the code were transported from specific
clusters to the general environment using .push and .pull commands
of the ipyparallel package.

\section{Results\label{sec:Results}}

\begin{figure*}
\begin{centering}
\includegraphics[angle=-90,width=0.9\textwidth]{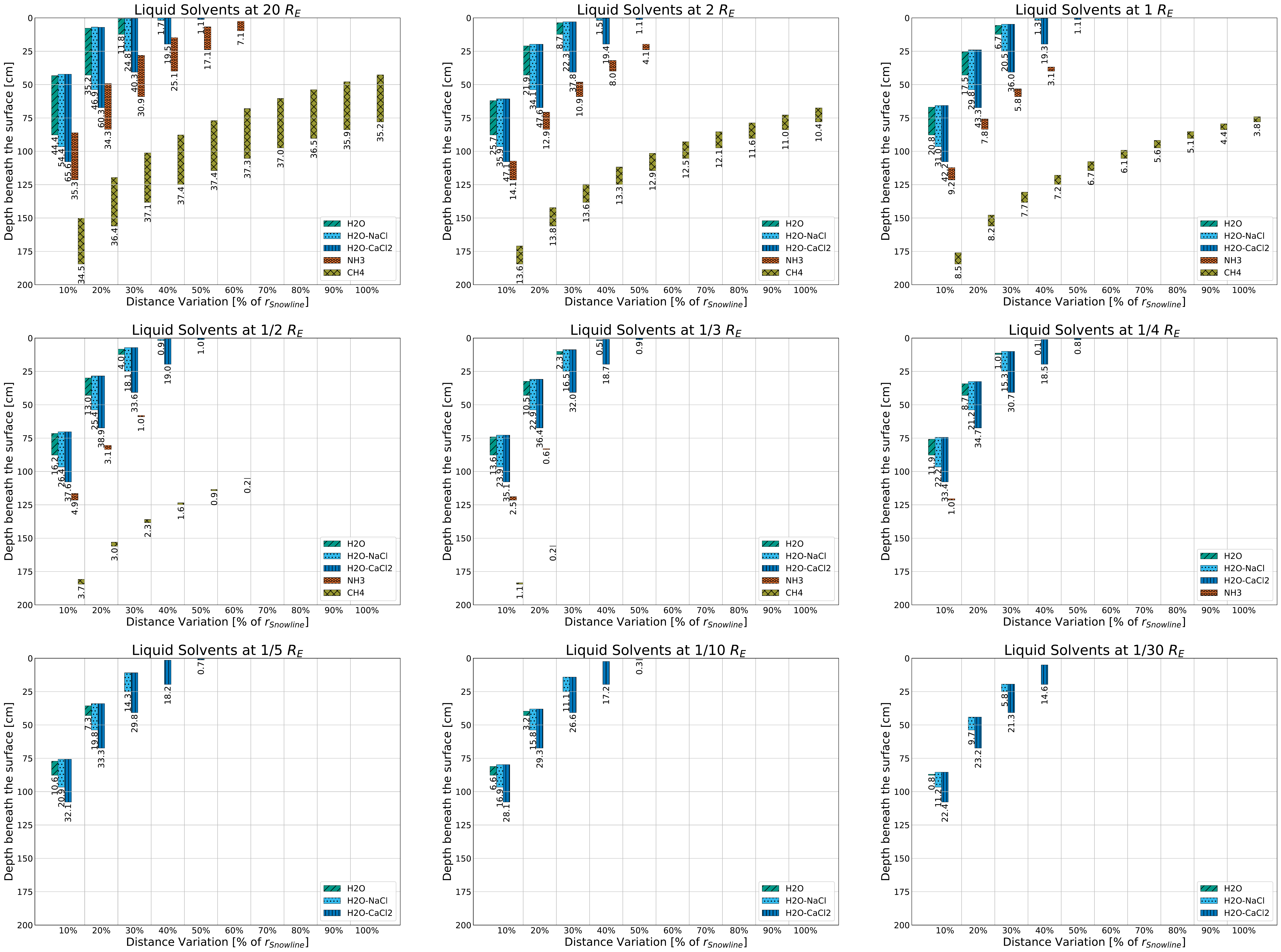}
\par\end{centering}
\caption{\label{fig:RockyFreshLiquid}Liquid solvent layers within the surface
of 9 model bodies of rocky, freshwater surface composition around
a modeled X-ray source.From top left to bottom right, the 9 bodies
are of the sizes $20,1,2,\frac{1}{2},\frac{1}{3},\frac{1}{4},\frac{1}{5},\frac{1}{10},\frac{1}{30}R_{earth}$respectively.
The surface density of the rocky, freshwater model involved is $\rho_{surface}=2.2765\frac{g}{cm{{}^3}}$,
with the composition shown in Table \ref{tab:Abundances}. Numbers
labelling the columns show the exact thickness of the liquid layers.}
\end{figure*}
The results displayed in Fig. \ref{fig:RockyFreshLiquid} are a detailed
breakdown of where, and under what conditions, the discussed solvents
would be liquid given the rocky, ``lunar soil'' model with freshwater
ices. (Results for the other surface models as well as some further
conditions considered can be found in Appendix \ref{sec:Exhaustive-model-simulations}.)
The 9 sub-plots seen show the nine sizes of model bodies considered
in this work, from largest at 20 earth radii to smallest at 1/30 earth
radii - smaller sizes are discussed in Appendices \ref{subsec:rhobodies}
and \ref{subsec:Determining-the-minimum}. The y-axes display model
depth beneath the surface, with the surface itself (depth 0) placed
at the top of each sub-plot. The x-axes display groupings of the ten
considered orbital distances between 10\% and 100\% $r_{snowline}$.
Within each of these groupings, the five different solvents considered
are displayed with different colors. As such, the bars display where
solvents would be liquid and are labeled with the respective bar's
length (representing the thickness of the liquid layer), taking both
the subterranean pressure and temperature into account. We can then
make some key observations based on these results:
\begin{enumerate}
\item It is possible for solvents to be liquid on (and in) bodies orbiting
an AGN at a few parsecs of distance, fueled entirely by the X-ray
emission of that AGN. This can happen in a somewhat wide variety of
environments, with the AGNs own energy output having much less of
an impact compared to the density of a surface regolith (which greatly
influences attenuation) or the size of the body itself (which greatly
influences phase changes). The body's inner density (beneath the regolith)
plays a vital role as well, as seen in Appendix \ref{subsec:rhobodies}.
The influences are stronger for $CH_{4}$and $NH_{3}$.
\item Salt content of optimal concentration has an immensely beneficial
effect on the thermal properties of water in extreme environments,
and one can easily imagine that even away from the optimal concentration
saltwater could beget liquidity on otherwise frozen worlds.
\item Unsurprisingly, different solvents are constrained by different variables
in this simulation, depending on the respective thermal properties.
Water's main constraint is temperature: Freezing points determine
if and at what distance from the central source liquid layers can
form, while boiling points mainly determine how deep within the crust
this has to happen. On bodies too close to the central source, water
close to the surface evaporates (the consequences of which would lead
to stark alterations in the model itself that have not been considered
here, as it would be beyond the scope of this paper) while water too
far below would not be warm enough to melt.Methane and ammonia are
constrained mainly by pressure, which allows these substances to be
liquid even close to the snowline if the pressure is sufficient -
so if the body is large (and dense) enough.
\end{enumerate}
To illustrate these key observations, we can determine that layers
of liquid fresh water inside a rocky, lunar-like model with the largest
thickness of $65$ to $44cm$ would be found between $20$ to $\frac{1}{5}R_{earth}$.
Salt water behaves similarly, but can also form thick (up to roughly
65 cm) liquid layers on even smaller bodies down to $\frac{1}{30}R_{earth}$thanks
to the low triple point pressure ($1mbar$ and $0.2mbar$ for optimal
concentrations of $NaCl$- and $CaCl_{2}$-saltwater respectively,
as opposed to $6.1mbar$ for freshwater).

The liquid layers may also be formed within bodies of lower densities
(as shown in Fig. \ref{fig:rhobodies}) or much smaller radii down
to just$\frac{1}{1250}R_{earth}$(as shown in Fig. \ref{fig:minR_rocky})
under certain conditions, especially for $CaCl_{2}$-saltwater.

A complete opposite behavior is seen within methane and ammonia: for
those, triple point pressures are high, making them unable to exist
in liquid form on bodies without a large radius in this model (and
a high mass in general). However, on large enough bodies, methane
and ammonia exhibit great versatility thanks to their temperature
indifference. As such, methane can persist in liquid form on bodies
at or above $\frac{1}{3}R_{earth}$(but mostly if close to the AGN
on the small bodies), and is able to cover the whole range of distances
between 10 and 100\% $r_{snow}$on bodies larger than 1$R_{earth}$.
Ammonia, while less versatile, can be liquid up to 60\% snowline distance
for $20R_{earth}$and is able to form a slim layer at 10\%$r_{snow}$
on bodies the size of $\frac{1}{4}R_{earth}$. The details of other
models and their respective constraints are expanded upon in the appendix.

$ $

Further variations of parameters and their influence on the stability
of the liquid systems were investigated by considering both the daylight
side of a possibly tidally locked body and limited the input to the
two extremes of the AGN sample (the strongest and weakest non-outlier
AGN). It can be shown that within the investigated group of AGN, there
is little difference between strongest, weakest and the averaged spectra
and the overall impact of the spectrum used as input is overshadowed
by the greater impact of the chosen surface and body models. On the
other hand, and as would be expected, the day-and-night cycle (and
lack thereof) have a far larger impact of the temperature and subsequently
liquid layers of solvents beneath a body's surface. A system without
a day-and-night cycle generally pushes liquid layers further into
the rock (however, we do not further inspect vaporization effects
that might occur on the surface under higher temperatures, lending
themselves to the formation of cryovolcanic activity) and allows for
liquid layers on distances further from the central source, where
liquid solvents where not possible on a rotating model. The detailed
breakdown of this can be seen in Appendix \ref{sec:Further-variabilities-investigat}.

\section{Discussion}

Given the wide slew of parameters tested (different model body sizes,
surface compositions and solvents considered), we consider it to be
fairly certain that liquid solvents are possible on bodies in such
an exotic environment as in an orbit around an active galactic nucleus.
In the same vein, this wide slew of parameters together with the current
lack of research on the topic does not allow a clear statement on
the propabilities involved, which was, however, not the point of this
work to begin with. It thus serves as a good base for further, even
more complex, models.

One aspect to investigate in such future models may be the matter
of crust stability. Several factors impact the stability of a celestial
body's crust (present liquids, the presence and - if so - nature of
an atmosphere and overall climate situation), and can do so in a drastic
manner. Just in the solar system, we know the moon's regolith is stable
enough for humans to comfortably walk on, while some research has
suggested the regolith of some icy moons such as Europa is so loose
that it would be more akin to light, fresh snow than actual solid
ground (\citealt{Nelson2016}). It would not be far from reason to
also expect a complex interplay of solid, gas and liquid phase to
result in unstable crusts in the form of cryovolcanism as expected
on Titan (\citealt{Mitri2008}), especially in the models showing
shallow water layers where eruption or refreezing events might lead
to significant pressure buildup (while such a risk seems less significant
for the often deeper layers of methane and ammonia). As a result of
this, despite what the wider liquid layers found on larger model bodies
might suggest, smaller model bodies are actually advantageous for
the existence of liquid ammonia and methane. These volatiles would
greatly destabilize the crust if too close to the surface, and smaller
bodies (with therefore less gravity at play) necessitate any potential
liquid layers to exist deeper beneath the surface, in turn stabilizing
it indirectly. Even harsher conditions arise for models very close
to the central source (at $\leq20\%\ r_{snowline}$), as temperature
profiles in these cases can be destructive (exceeding the melting
points of rocky compounds), endangering the stability of even rocky
crusts and necessarily needing evaporative effects of involved volatiles
to be taken into account for a full assessment of the situation. As
such, any results from $10\%$ to $20\%\,r_{snow}$have to be investigated
cautiously and should be reconsidered in new, more refined models.
Similar considerations may be true for icy crusts close to the snowline.
We also expect the rotation of a model body to play a large role,
with the effects of a tidal lock having been shown here already. It
can be expected that the opposite, a body with an anomalously high
rotational velocity, would suffer problems of its own. These outliers
would thusly be perfect candidates for exogeological and exoplanetary
investigations of expectable surface composition and structure of
a circumnuclear body.

This would also allow for more precise considerations on body size,
a topic that has only been generally outlined in this work. While
we find liquid methane and ammonia on bodies larger than half the
radius of earth (equivalent to the Jovian moon Ganymede) at a depth
of around 1 meter, liquid fresh and saltwater layers can effectively
be found in bodies as small as $\frac{1}{30}R_{earth}$ (equivalent
to the range of large asteroids such as Vesta), with both $NaCl$
as well as $CaCl_{2}$ proving beneficial to a liquid phase. $CaCl_{2}$-saltwater
alone may even be found in liquid layers on bodies ranging down to
$\frac{1}{1250}R_{earth}$(equivalent to the realm of comets under
special conditions), as shown in Fig. \ref{fig:minR_rocky}. On the
other end of the scale, large bodies would pose their own set of challenges
and interesting aspects that might influence these results, such as
an accumulation of radioactive material and stored accretion energy
begetting internal heating (similar to processes inside earth) or
perhaps even the formation of gas giants under certain conditions.

On the topic of the aforementioned outliers, we lastly want to mention
that especially the thermal buildup-effects (that result from hardcoding
the equilibrium temperature at the outermost surface layer into the
model), which can be seen in Appendix \ref{sec:Exhaustive-model-simulations},
make it apparent that the thermal model employed here has very apparent
limits. 2D heat transfer effects and, most notably, the tremendous
impact of an atmosphere are also topic for further research. All of
these open up possibility for an improved thermal model in future
works.

These prospects about liquid solvents of course serve the greater
purpose of enabling us to make educated considerations about the possibility
of life in such exotic environments as seen here. Liquid layers of
thicknesses between half a meter and a few millimeters seem unconventionally
small biospheres, but similar microhabitats have already been considered
on earth (\citealt{Fritsen2001}, \citealt{Glass2021}) and continue
to grow in relevancy for astrobiology. This then warrants thoughts
about the usability (and hostility) of the energy not just for the
potential habitat itself, but the organisms within it as well. Models
for the conversion of gamma-ray energy into biomass have already been
proposed (\citealt{Altair2018}, \citealt{Braun2016}) before. But
these models were developed for or from low energy environments, which
makes it questionable if or how they may be applied to high energy
environments, and applications of these models here would lead one
to expect high, probably unrealistic amounts of potential biomass.
Especially given that new restrictions on growth and homoeostasis
by the amount of gamma-radiation, the emergence and amount of secondary
and tertiary particles, or potential new mechanisms for converting
energy into useable units for living entities may play a bigger role.
We would believe that a thoughtful analysis of the possibilities of
gamma-ray driven life in this environment may deserve much deeper
investiagtion, and as such consider that beyond the scope of this
paper.

Our aim with this work was testing the waters (quite literally so)
of a rather exotic, high-energy environment, and to add to the increasing
awareness of the potential that circumnuclear clouds around active
galactic nuclei hold. Our results show that solvents in such a model
can exist in liquid phase even over a wide set of parameters. We thusly
point to a number of new channels, especially in geology and astrobiology,
for both in-depth theoretical work surrounding this and similar topics,
as well as experimental investigations (such as remote sensing of
appropriate sources or laboratory models of such environments) and
deem our initial goal as succesfully reached.
\begin{acknowledgements}
I would like to thank Malice Rudolph for aid in optimizing accessibility
of the core figures in this work, as well as continued support concerning
the English language.
\end{acknowledgements}

\bibliographystyle{aa}
\nocite{*}
\bibliography{manuscriptbiblio}

\begin{appendix}

\section{Obtaining distances to investigated AGNs using astroquery and SIMBAD\label{sec:Obtaining-distances-to}}

To determine the flux arriving at a model planet located a certain
distance from the central object, we first need to correct measurements
taken for distance and viewing angle from both the measurements and
the model planet in relation to the object's geometry. To apply the
former using the inverse square law as will be shown in (\ref{eq:DistAdjust})
we need distances from earth to the different galaxies of the sample.

Where possible for each of these objects, a direct luminosity distance
was obtained from the SIMBAD online library. Where this was not possible,
a redshift z was still available from SIMBAD and as a consequence
the estimated distance D was approximated using Hubble\textquoteright s
law:

\noindent
\begin{equation}
D\approx\frac{cz}{H_{0}},\label{eq:A}
\end{equation}

\noindent with $c$ being the velocity of light in vacuum, and using
a hubble parameter of $H_{0}=69.9\frac{km}{s\,Mpc}$ (\citealt{Khetan2021}).

Accessing SIMBAD was done via the astroquery python package (\citealt{Astroquery}),
specifically astroquery.simbad, and setting up a custom query using
object denominators taken from the filenames of (\citealt{AGNSEDATLAS})-data.

\section{Exhaustive model simulations and temperature curves\label{sec:Exhaustive-model-simulations}}

The main product of the actual thermal simulation is a temperature
profile of the subsurface environment considered, as they preceed
the liquid layers determined as shown in section \ref{sec:Results}.

\begin{figure}[h]
\centering{}\includegraphics[width=1\columnwidth]{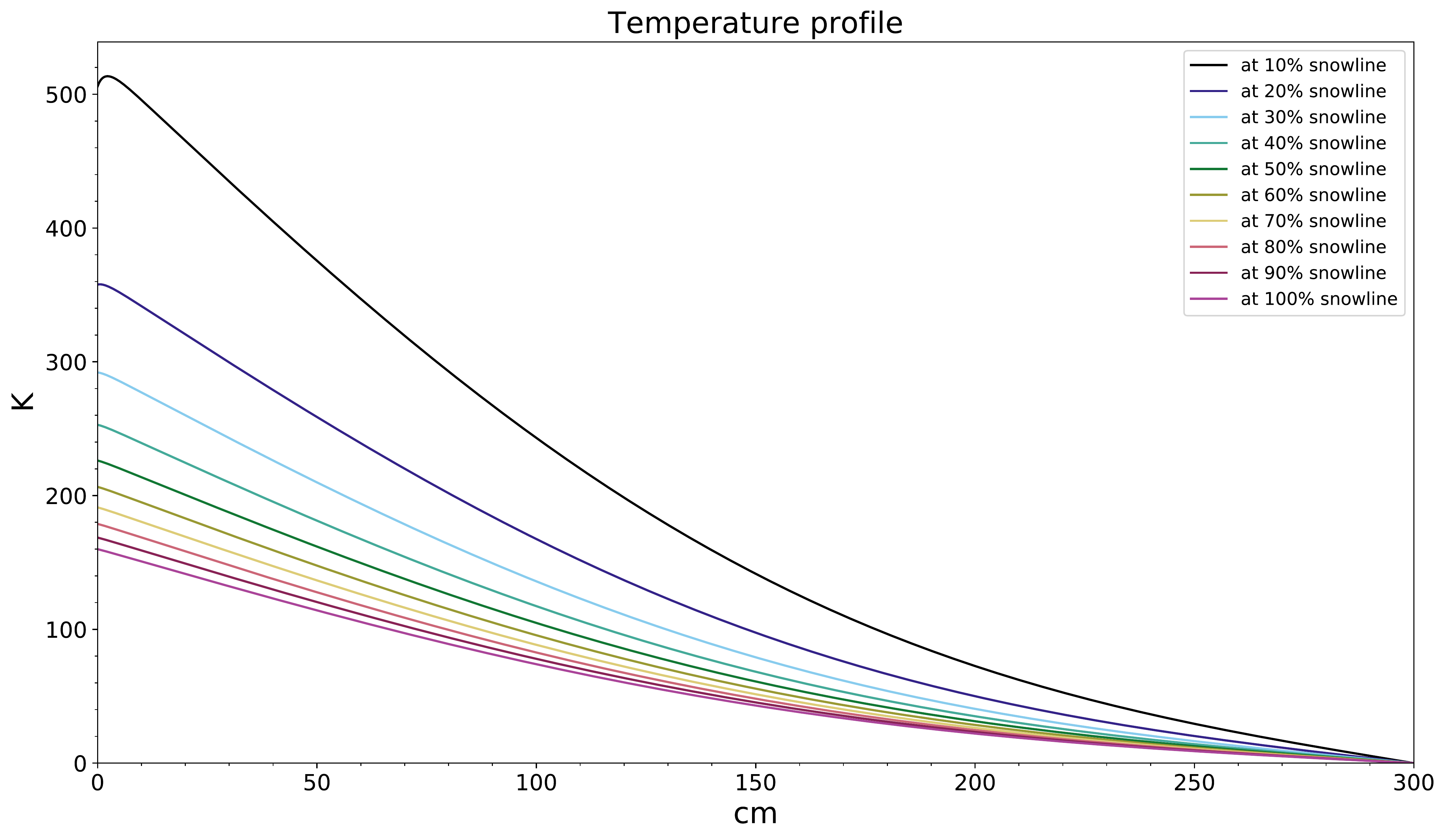}\caption{\label{fig:RockyFreshwaterProfile}Temperature profile for the simulation
using a rocky crust with freshwater inclusions.}
\end{figure}

\begin{figure*}
\subfloat[Ice, massive]{\begin{centering}
\includegraphics[width=0.5\textwidth]{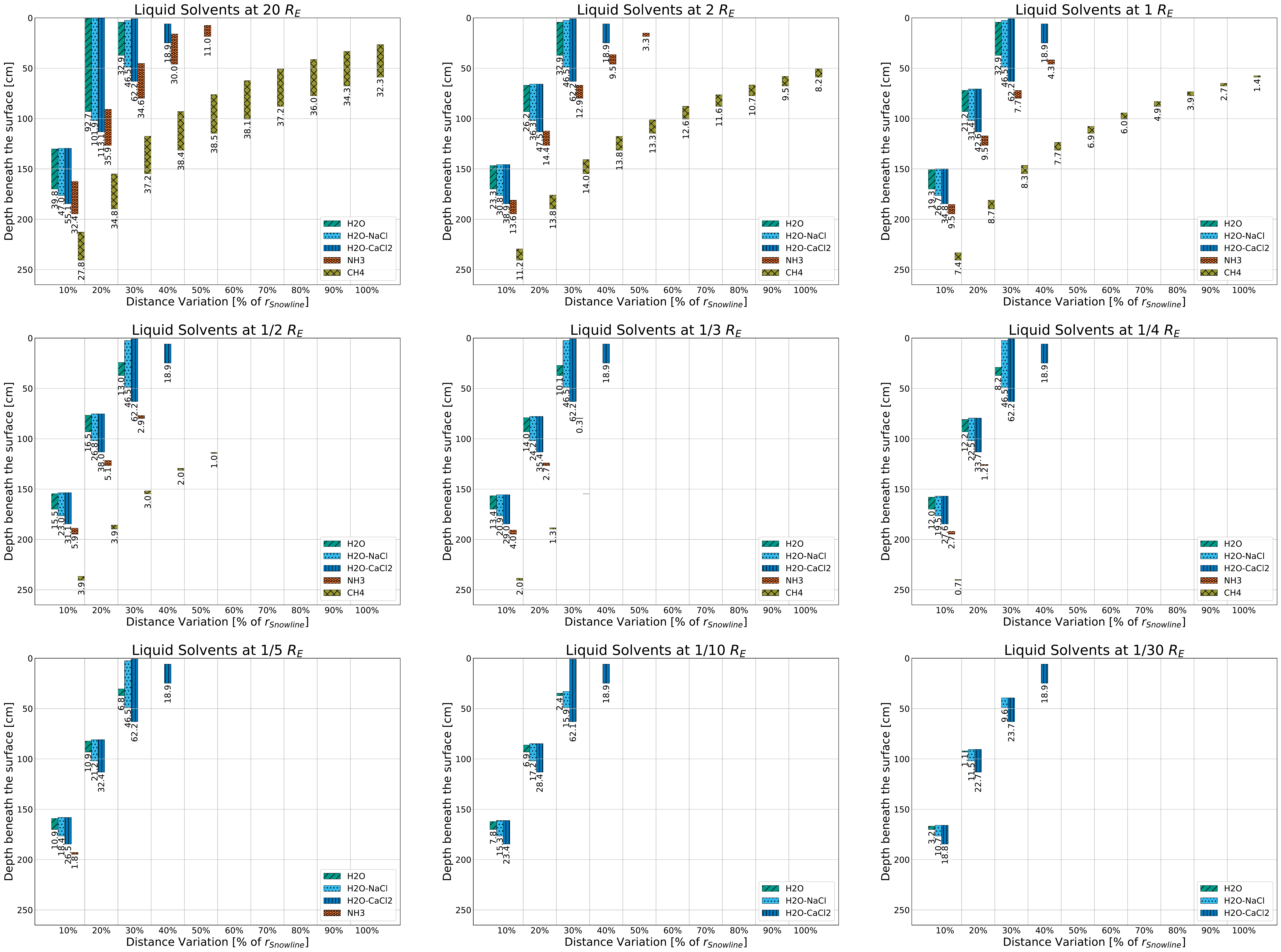}\includegraphics[width=0.5\textwidth]{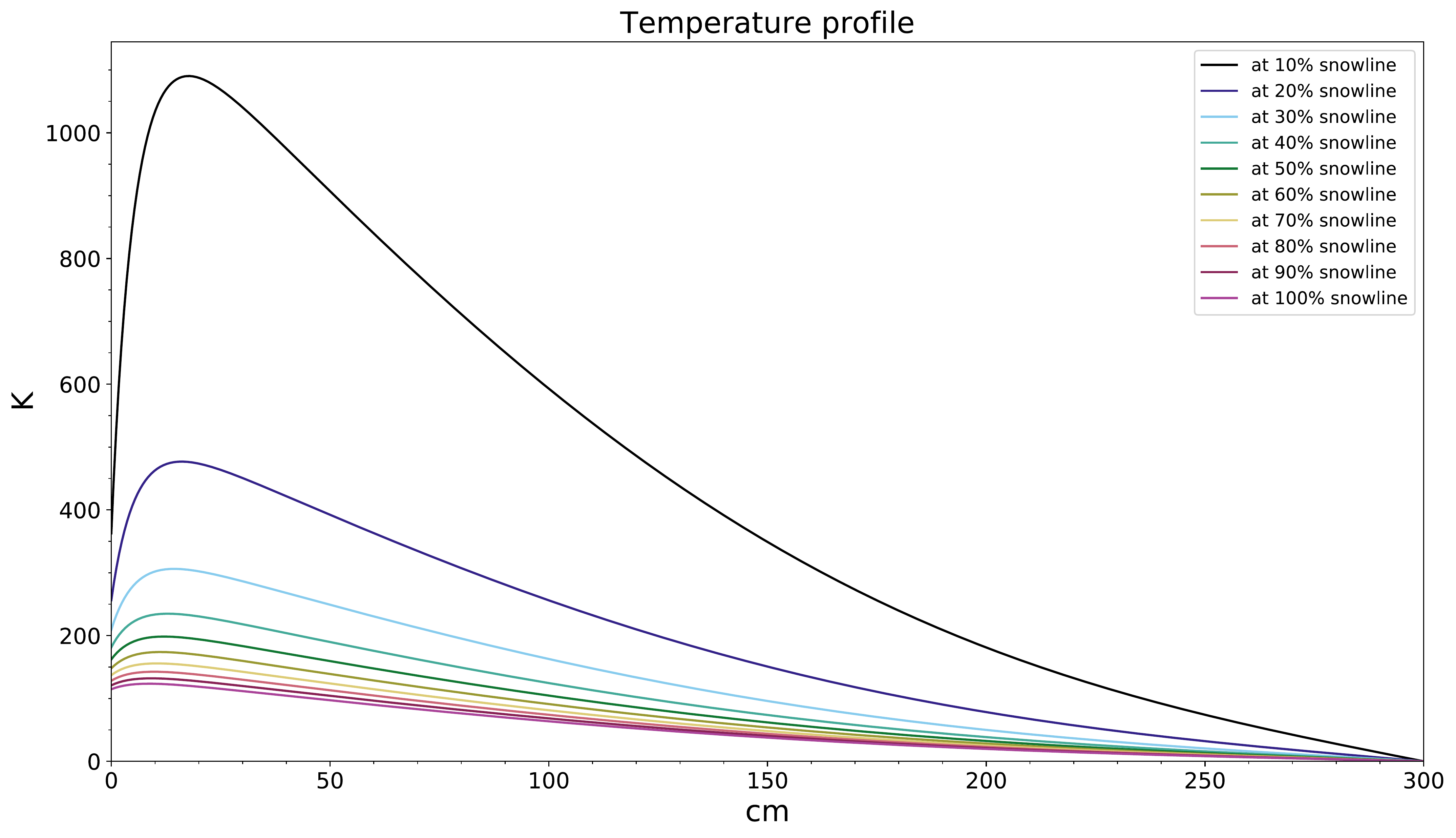}
\par\end{centering}
}

\subfloat[Rocky, $NaCl$-saltwater ]{\begin{centering}
\includegraphics[width=0.5\textwidth]{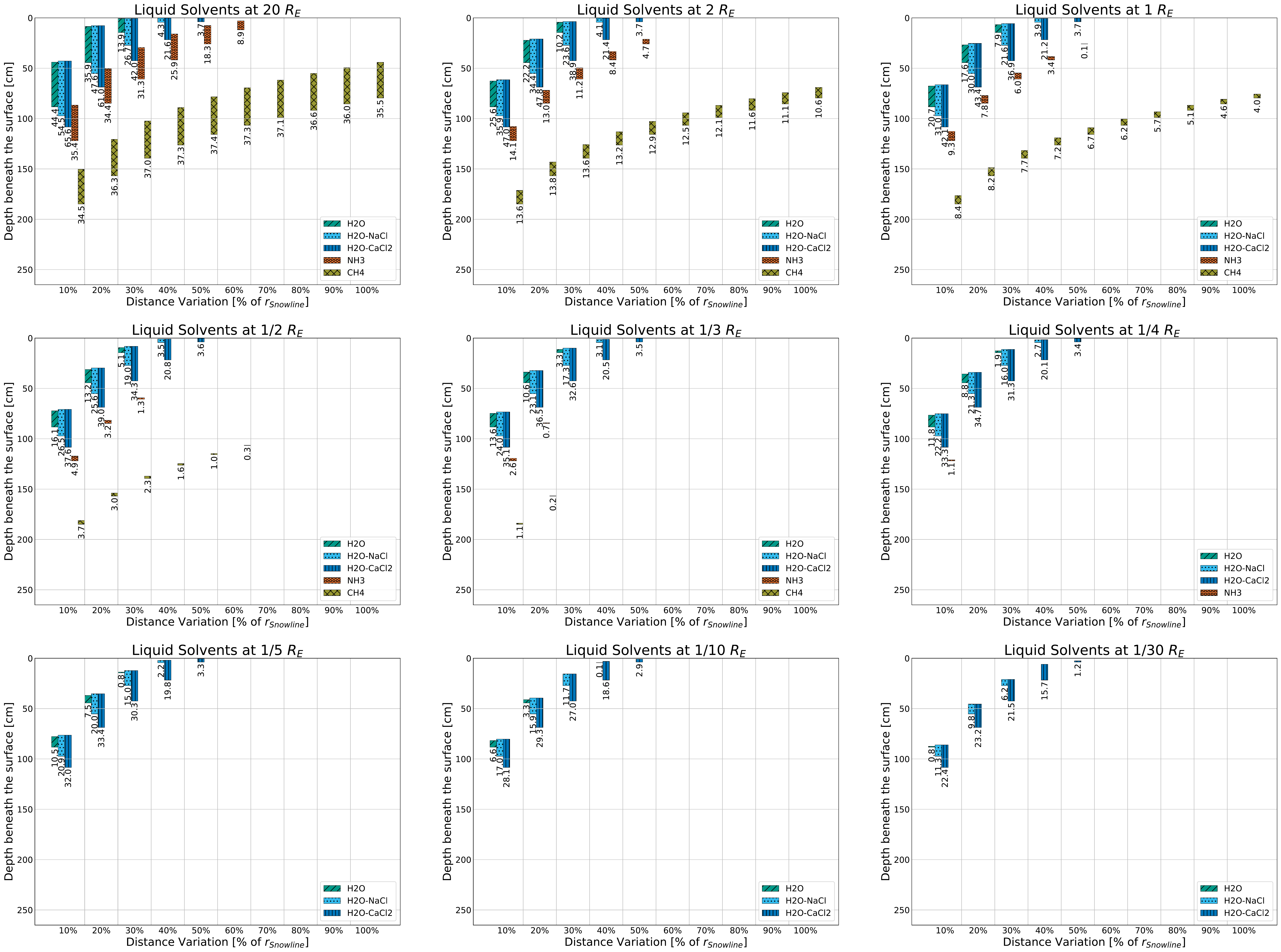}\includegraphics[width=0.5\textwidth]{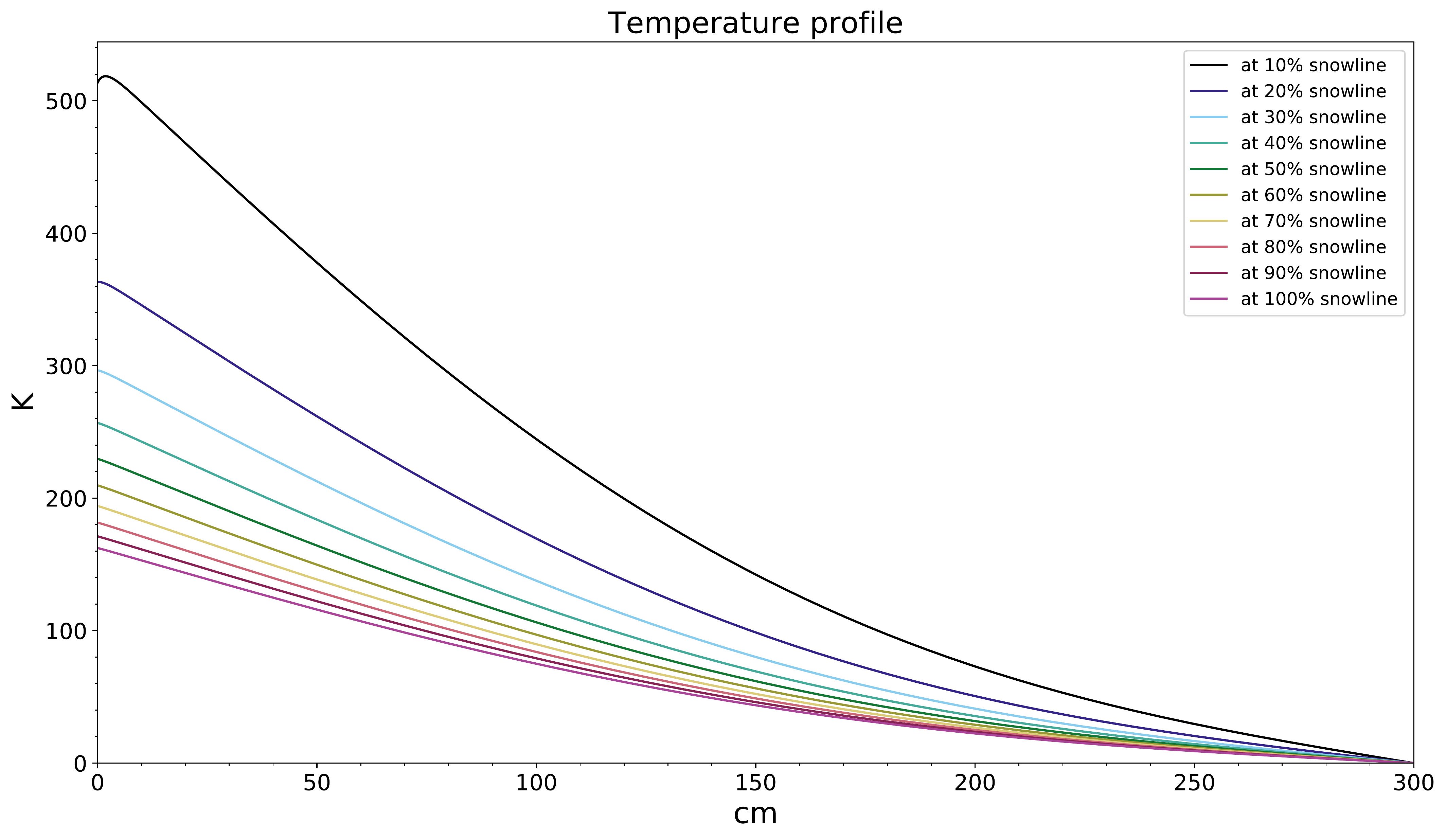}
\par\end{centering}
}

\subfloat[Rocky, $CaCl_{2}$-saltwater]{\begin{centering}
\includegraphics[width=0.5\textwidth]{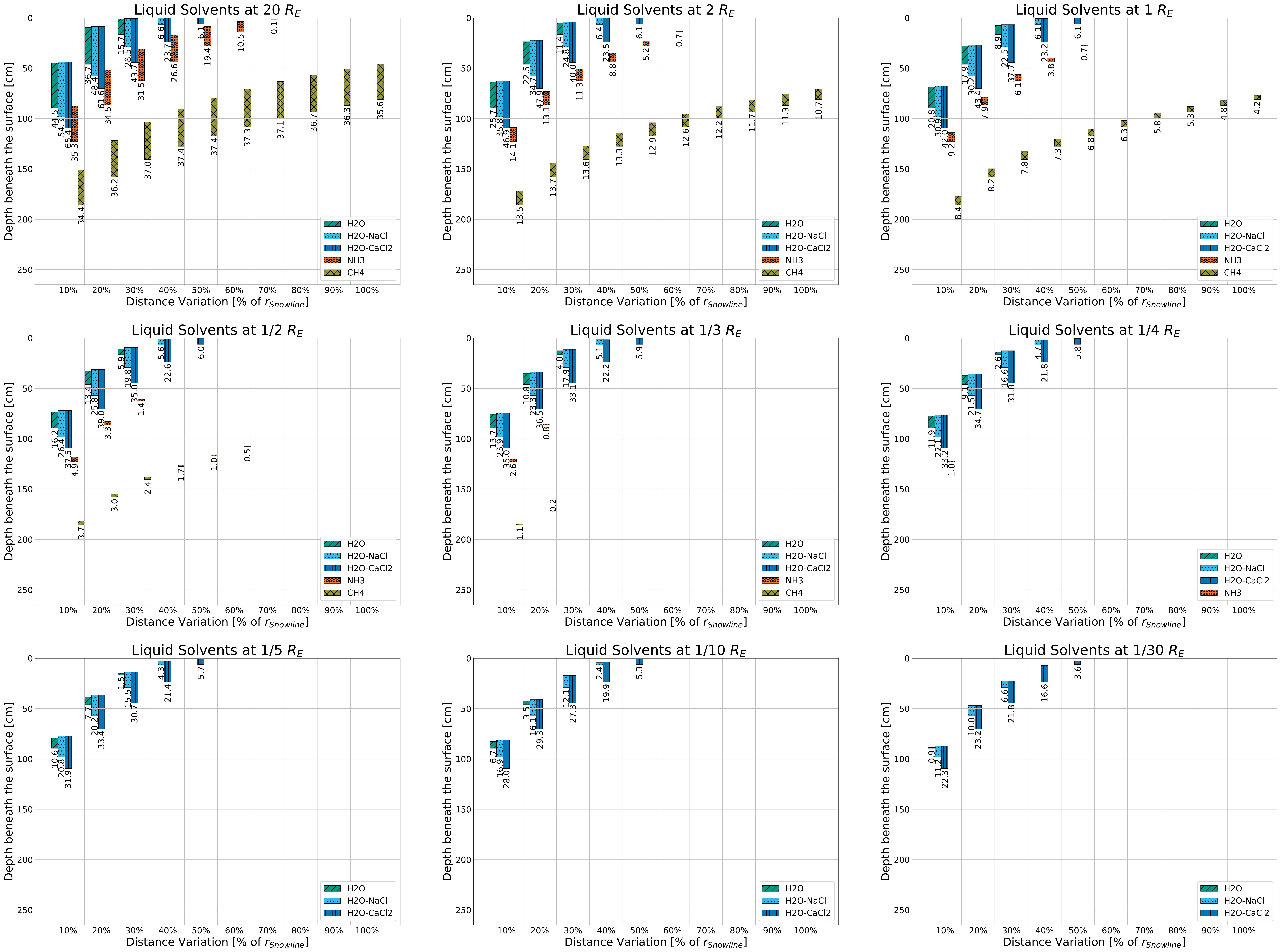}\includegraphics[width=0.5\textwidth]{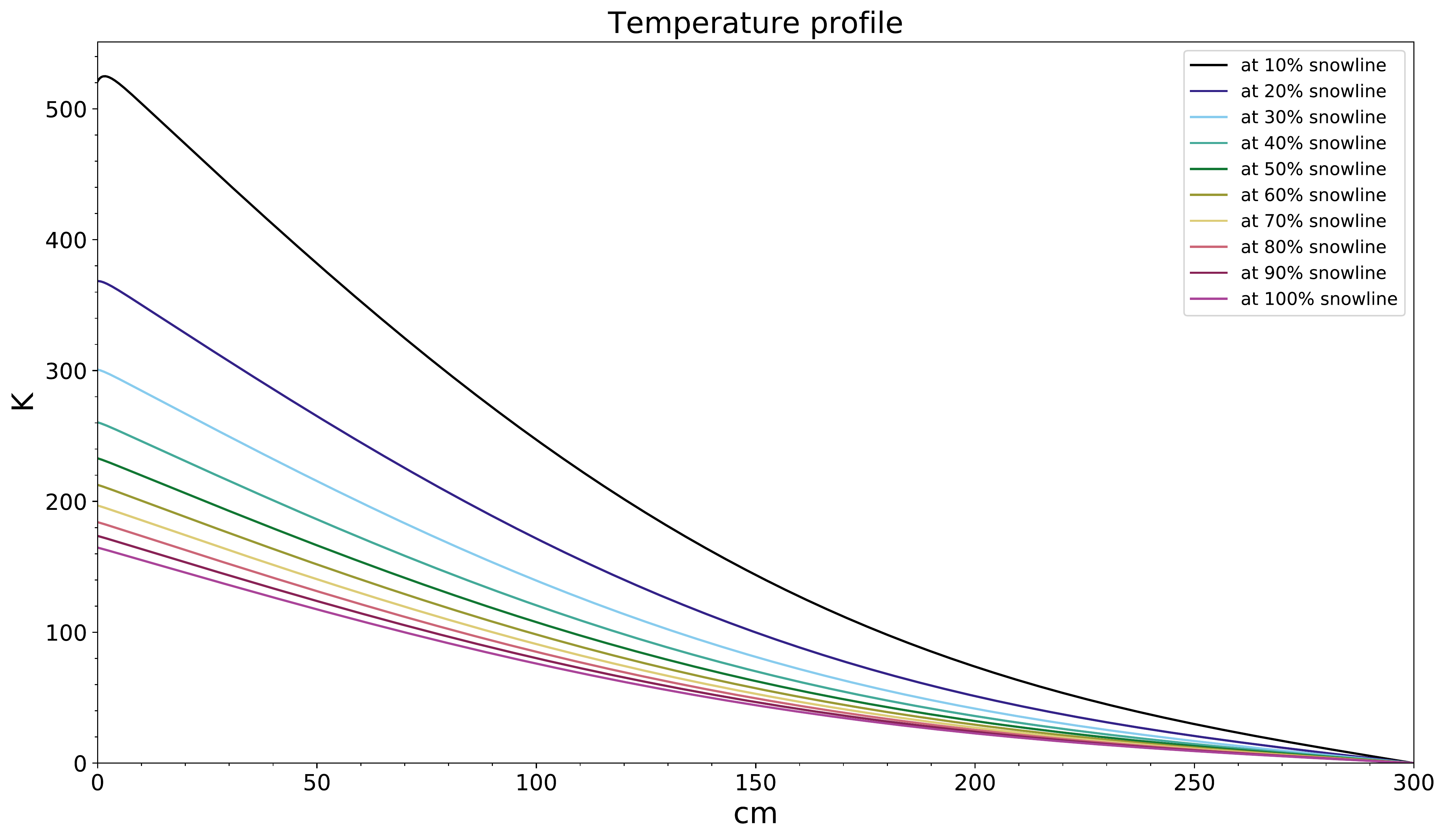}
\par\end{centering}
}\caption{\label{fig:ExhaustSims}Simulations for the three exhaustive simulations
of rocky crust with NaCl-saltwater inclusions, rocky crust with CaCl2-saltwater
inclusions and icy crust}
\end{figure*}

To keep the main part of this work focused, we show these temperature
profiles here, alongside outcomes of the previously mentioned additional
models, in Fig. \ref{fig:ExhaustSims} (apart from the main model
pertaining a rocky, freshwater-filled crust, the temperature profile
of which can be seen in Fig. \ref{fig:RockyFreshwaterProfile}).

\section{Mass attenuation coefficient\label{sec:Mass-attenuation-coefficient}}

\begin{figure*}
\centering{}\includegraphics[width=1\textwidth]{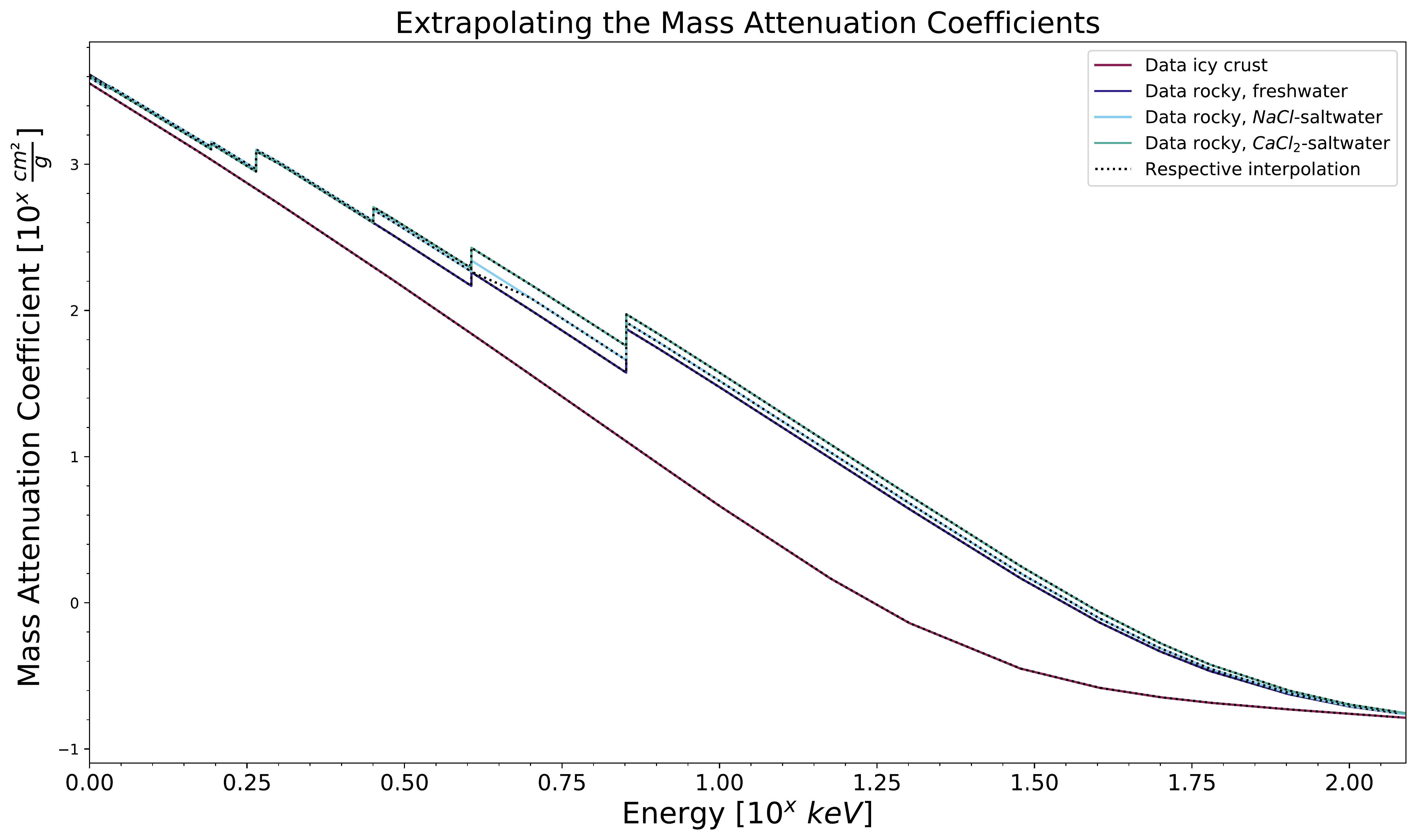}\caption{\label{fig:MassAttCoef}Mass attenuation coefficient for four surface
compositions.}
\end{figure*}

The mass attenuation coefficients for the considered surface compositions,
interpolated from data obtained from the NIST \citeauthor{NISTcalc}
database as in explained in Section \ref{subsec:Calculating-attenuated-energy}
is shown in Fig. \ref{fig:MassAttCoef}.

\section{About heat conversion\label{sec:About-heat-conversion}}

As mentioned, when radiation hits a material not all of the energy
is converted purely into heat with some energy resulting in secondary
radiation or the chemical ``upgrade'' (breaking and reformation
of chemical bonds, resulting in new compounds with the potential to
store more energy than before) of the matter hit by radiation. While
these secondary processes lie beyond the scope of this work, we wish
to make a conservative estimation on the amount of photon energy that
actually gets converted into heat to adjust the thermal model therewith.

This estimation is based on evaluating the amount of ``pure'' kinetic
energy released in the medium as minimum input for heating (secondary
and tertiary particles will also still render part of their energy
into heat). This was done utilizing the mass energy-absorption coefficient
$\frac{\mu_{en}}{\rho}$, which expresses the amount of energy from
the incident photon that is transferred as kinetic energy to charged
particles in the interaction minus the energy from photons resulting
from the movement of these charged particles. When compared to the
mass attenuation coefficient $\frac{\mu}{\rho}$, this lets us estimate
the portion of attenuated energy ``lost'' to heating.

For a variety of substances (comparable to the compounds involved
in our model, as no exact data for those could be obtained), in the
energy range with the most impact from the analysed AGN spectra (50
keV to 150 keV) and for densities $\rho=1\frac{g}{cm{{}^3}}$more
than 80\% and less than 90\% of energy is absorbed via the mechanisms
considered under$\frac{\mu_{en}}{\rho}$ (see Table \ref{tab:EnergyTransferCoef}).
At higher densities as they would be seen in the rocky compositions,
absorption reaches more than 90\% across the board.

This further encourages us to choose 0,9 as a very conservative factor,
posing a general minimum of energy available to the thermal model.\textquotedbl{}

\begin{table}[H]
\caption{\label{tab:EnergyTransferCoef}Mass energy absorption and mass attenuation
coefficients.}

\begin{centering}
\begin{tabular}{l|c|c|c|c|c}
\hline
Substance & \multicolumn{5}{c}{$E\ [keV]$}\tabularnewline
 & 50 & 60 & 80 & 100 & 150\tabularnewline
\hline
\multicolumn{1}{l}{for $\rho=1\frac{g}{cm{{}^3}}$ } & \multicolumn{5}{c}{$\exp(\frac{\mu_{en}}{\rho}-\frac{\mu}{\rho})\ [\%]$}\tabularnewline
\hline
water, liquid & 83.14 & 84.03 & 85.41 & 86.48 & 88.44\tabularnewline
glass & 83.64 & 84.77 & 86.36 & 87.49 & 89.44\tabularnewline
concrete & 83.12 & 84.33 & 85.99 & 87.17 & 89.17\tabularnewline
\hline
\multicolumn{1}{l}{for $\rho=2\frac{g}{cm{{}^3}}$ } & \multicolumn{5}{l}{}\tabularnewline
\hline
water, liquid & 91.18 & 91.67 & 92.42 & 93.00 & 94.04\tabularnewline
glass & 91.45 & 92.07 & 92.93 & 93.54 & 94.57\tabularnewline
concrete & 91.17 & 91.83 & 92.73 & 93.36 & 94.43\tabularnewline
\hline
\multicolumn{1}{l}{for $\rho=3\frac{g}{cm{{}^3}}$} & \multicolumn{1}{c}{} & \multicolumn{1}{c}{} & \multicolumn{1}{c}{} & \multicolumn{1}{c}{} & \tabularnewline
\hline
water, liquid & 94.03 & 94.36 & 94.88 & 95.27 & 95.99\tabularnewline
glass & 94.22 & 94.64 & 95.23 & 95.64 & 96.35\tabularnewline
concrete & 94.02 & 94.48 & 95.09 & 95.53 & 96.25\tabularnewline
\hline
\end{tabular}
\par\end{centering}
A list of mass energy absorption coefficient $\frac{\mu_{en}}{\rho}$
values, mass attenuation coefficient $\frac{\mu}{\rho}$ values, both
from (\citealt{NISTMassAbsorption}), and the portion of attenuated
energy NOT lost in secondary processes $\exp(\frac{\mu_{en}}{\rho}-\frac{\mu}{\rho})$.
\end{table}

\section{About the triple point (equivalent)\label{sec:About-the-triple}}

\subsection{Triple point (equivalent) properties}

The properties for fresh water, methane and ammonia have been obtained
from a NIST database. The matter gets more complicated when dealing
with saltwater however, it can be shown (\citealt{Bodnar2001}), that
the triple point equivalent\footnote{Triple points are strictly speaking only defined for pure substances,
we here use the critical point equivalent of a triple point as we
are not interested in the exact phase behavior, merely the minimum
pressure and temperature necessary for liquids to occur.} of saltwater solutions at their eutectic points can be approximated
by the vapoure pressure of pure water ice at that temperature. So,
using the known temperatures of the eutectic points of $NaCl$-Saltwater
at $23wt\%$ and $CaCl_{2}$-Saltwater at $30wt\%$ (\citealt{Ketcham1996})
we can determine the respective equivalent for the triple point pressure
from the behavior of fresh water. All resulting triple point (equivalents)
are shown in Table \ref{tab:TriplePointProperties}.

\begin{table}[H]
\caption{\label{tab:TriplePointProperties}Triple point (equivalent) properties}

\begin{centering}
\begin{tabular}{lll}
\hline
Formula & \multicolumn{2}{l}{Properties}\tabularnewline
\hline
 & $p_{triple}[bar]$ & $T_{triple}[Kelvin]$\tabularnewline
\hline
$H_{2}O$ & $0.0061^{\textrm{a}}$ & $273^{\textrm{a}}$\tabularnewline
$CH_{4}$ & $0.117^{\textrm{a}}$ & $91^{\textrm{a}}$\tabularnewline
$NH_{3}$ & $0.061^{\textrm{a}}$ & $195^{\textrm{a}}$\tabularnewline
\hline
$NaCl$ & $0.001^{\textrm{b}}$ & $252^{\textrm{b}}$\tabularnewline
$CaCl_{2}$ & $0.0002^{\textrm{b}}$ & $223^{\textrm{b}}$\tabularnewline
\hline
\end{tabular}
\par\end{centering}
$^{\textrm{a}}$NIST Chemistry WebBook, NIST Standard Reference Database
Number 69 (\citealt{NISTWebbook})

$^{\textrm{b}}$\citealt{Ketcham1996}
\end{table}

\subsection{Derivation of triple point depth\label{subsec:Derivation-of-triple}}

In equation (\ref{eq:TripleDepth}) the depth at which the triple
point pressure, that is the pressure necessary for the relevant solvents
to melt (instead of just sublimate), is calculated. The process of
arriving at this equation is outlined here:

The triple point pressure $p_{triple}$ is a material constant and
can be obtained from databases or literature, and can be expressed
as the gravitative force $F_{g}$ of (on an atmosphereless body: just)
the ground above an observed area $A$, which here lies at the desired
depth $d_{triple}$:

\begin{equation}
p_{triple}=\frac{F_{g}}{A},\label{eq:ptriple}
\end{equation}
The force can then be broken down, where in this case $m_{body}$
denotes the mass of the planet below the area at $d_{triple}$and
$m_{surface}$ denotes the mass of the crust weighing down from above:
\begin{equation}
F_{g}=G\frac{m_{body}m_{surface}}{r^{2}},\label{eq:graviforce}
\end{equation}
Since we later want to vary the object radius $R_{body}$ of hypothetical
bodies, but not their density $\rho_{body}$, which we set to $\rho_{body}=\rho_{earth,mean}=5.51\frac{g}{cm{{}^3}}$(\citealt{EarthFactSheet}),
to simplify the model, it is a good choice to further break down:

\begin{align}
m_{body} & =V_{body}\rho_{body}=\frac{4}{3}\pi R_{body}^{3}\rho_{body},\label{eq:mbody}\\
m_{surface} & =V_{surface}\rho_{surface}=Ad_{triple}\rho_{surface},\label{eq:msurface}
\end{align}
As our model in general only considers depths of up to 3 meters, it
is neglectable when compared to the radius of all the considered models
(between $20$ and $\frac{1}{30}R_{earth}$) so the radius in (\ref{eq:graviforce})
can be assumed to be equal to $R_{body}$. As such, we can insert
(\ref{eq:msurface}) and (\ref{eq:mbody}) into (\ref{eq:graviforce})
and this then into (\ref{eq:ptriple}) and simplify:

\begin{align}
p_{triple} & =\frac{1}{A}G\frac{\frac{4}{3}\pi R_{body}^{3}\rho_{body}Ad_{triple}\rho_{surface}}{R^{2}}\\
 & =G\frac{4}{3}\pi R_{body}\rho_{body}d_{triple}\rho_{surface},
\end{align}
which we can rearrange to reach our desired equation (\ref{eq:TripleDepth})
for the triple depth.

\section{Further variabilities investigated\label{sec:Further-variabilities-investigat}}

As a final step, we consider two points of variability not touched
upon before: one is the impact of AGN-flux, and by extension the mean
done to simplify calculations here, on temperature curve and liquidity
of solvents. The other is the impact of the implemented day-and-night-cycle
on the results.

\subsection{Simulation on a tidally locked body}

\begin{figure*}[p]
\caption{\label{fig:DaynNite}\protect\includegraphics[width=0.85\textwidth]{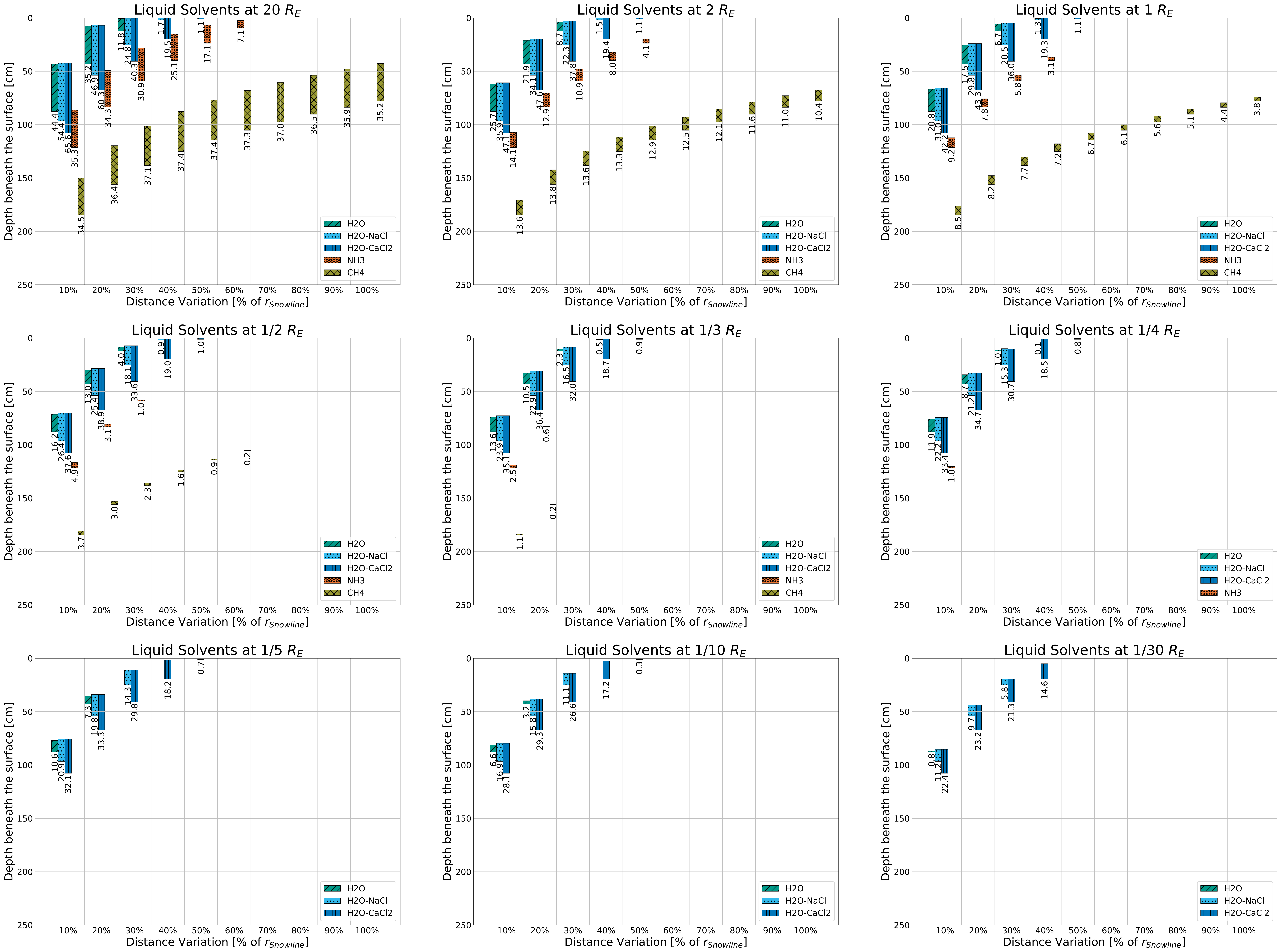}}

\caption{\label{fig:Allday}\protect\includegraphics[width=0.85\textwidth]{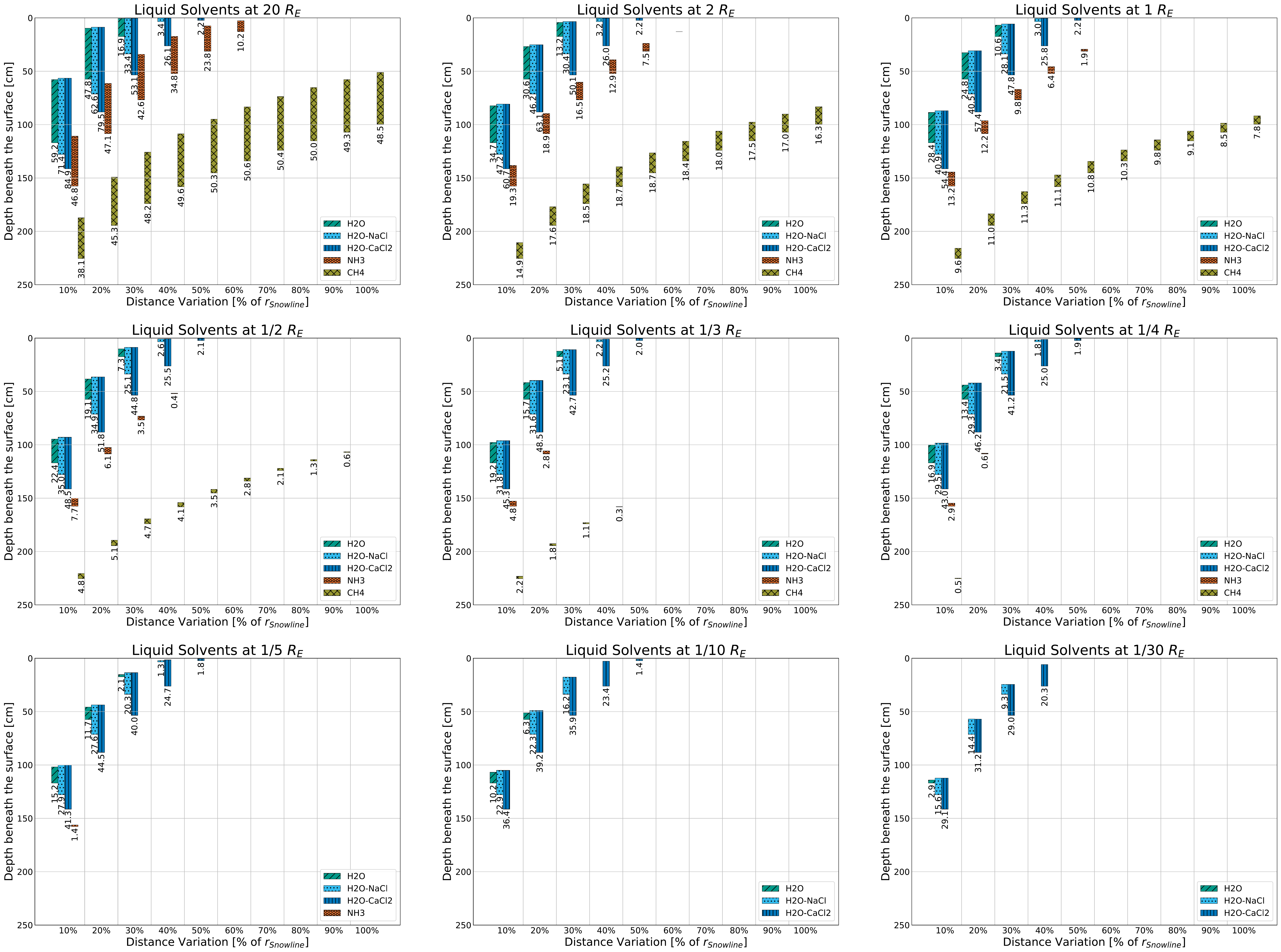}}
\ref{fig:DaynNite} Simulation of the rocky freshwater model with
a standard day-and-night cycle (same as Fig. \ref{fig:RockyFreshLiquid}).

\ref{fig:Allday} Simulation of the rocky freshwater model with a
continuous ``day'' as if tidally locked.
\end{figure*}

For this, we disable the part of the code that ensures the continuous
source term is added only every other timestep, simulating the lit
side of a body tidally locked to the AGN. (Terminator results are
not simulated, rotation is generally half the stepping.) While it
should be noted that tidal locking seems unlikely for bodies orbiting
several parsecs away from the source (even if said source is supermassive),
it allows us to investigate an extreme case. This will be especially
interesting in potential follow-up work considered atmospheric bodies.
Results, as can be seen in Fig. \ref{fig:Allday}b) for the rocky
freshwater model, while not extremely different to the normal simulation
(which is shown, with adapted scales, in Fig. \ref{fig:DaynNite}),
show obvious distinctions, especially that a ``tidally locked''
model allows liquid layers where none were possible on a rotating
body, most notably ammonia on bodies as small as $\frac{1}{5}R_{earth}$if
close enough to the source, or methane on bodies as far away from
the source as 40\% $r_{snow}$on bodies the size of $\frac{1}{5}R_{earth}$.

\subsection{Simulation with individual galaxies}

\begin{figure*}[p]
\caption{\label{fig:SpecgalWeak}\protect\includegraphics[width=0.85\textwidth]{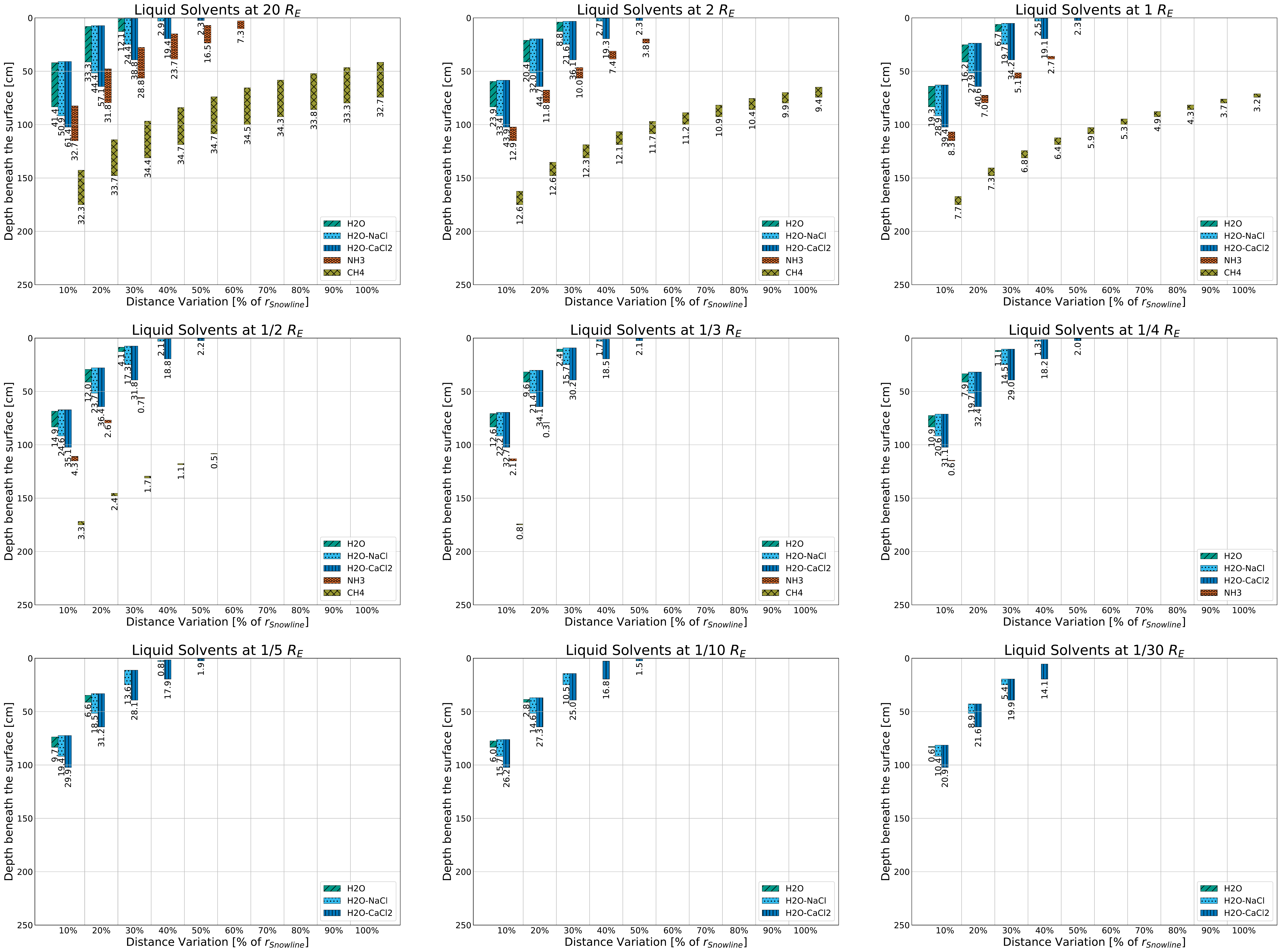}}
\caption{\label{fig:SpecgalStrong}\protect\includegraphics[width=0.85\textwidth]{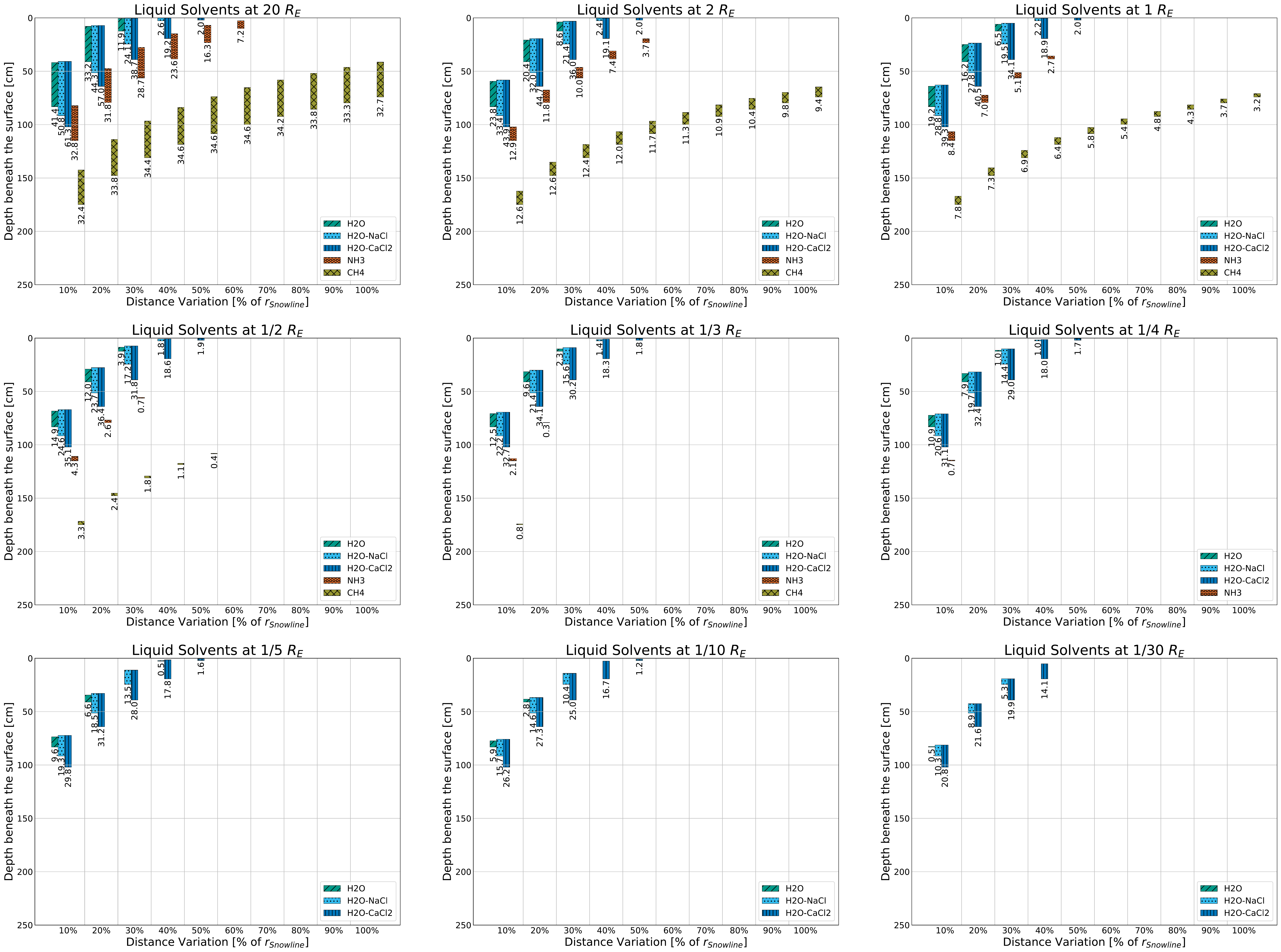}}

\ref{fig:SpecgalWeak} Simulation with input coming from a single,
weak source (NGC 3516), $r_{snow}=2.95pc$.

\ref{fig:SpecgalStrong} Simulation with input coming from a single,
strong source (Mrk 876), $r_{snow}=14.92pc$.
\end{figure*}

To compare the different impacts that strong and weak sources have
on the eventual results, two runs of the simulation were performed
not using the input of the averaged spectrum of all 20 AGN, but instead
spectra of two non-averaged, individual AGN. Specifically NGC 3516
and Mrk 876, which were identified as strongest and weakest non-outlier
source of the dataset respectively. The results, as can be seen in
Fig. \ref{fig:SpecgalWeak}and \ref{fig:SpecgalStrong}, differ only
barely from that of the averaged spectrum used for the main simulation,
which confirms that in this simulation, the strength of the source
(as long as said source is of the same, Seyfert 1, object class) has
only marginal impact on the temperature profile resulting from them.

What can also be seen is that counterintuitively, not only does the
weaker AGN result in wider liquid layers than the stronger one, but
both are outdone by the meaned input. The key factor at play here
is the spectrum's shape, namely their comparative flux at the highest
energies investigated. When normed for the same integrated flux (which
is the case for the flux at snowline), Mrk 876 contributes less high-energy
radiation compared to NGC 3516 (and both less compared to the meaned
spectrum), and as a result less radiation penetrates as deeply and
liquid layers are (ever so slightly) slimmer.

\subsection{Simulations of small moons and asteroids\label{subsec:rhobodies}}

\begin{figure*}
\begin{centering}
\includegraphics[width=0.9\textwidth]{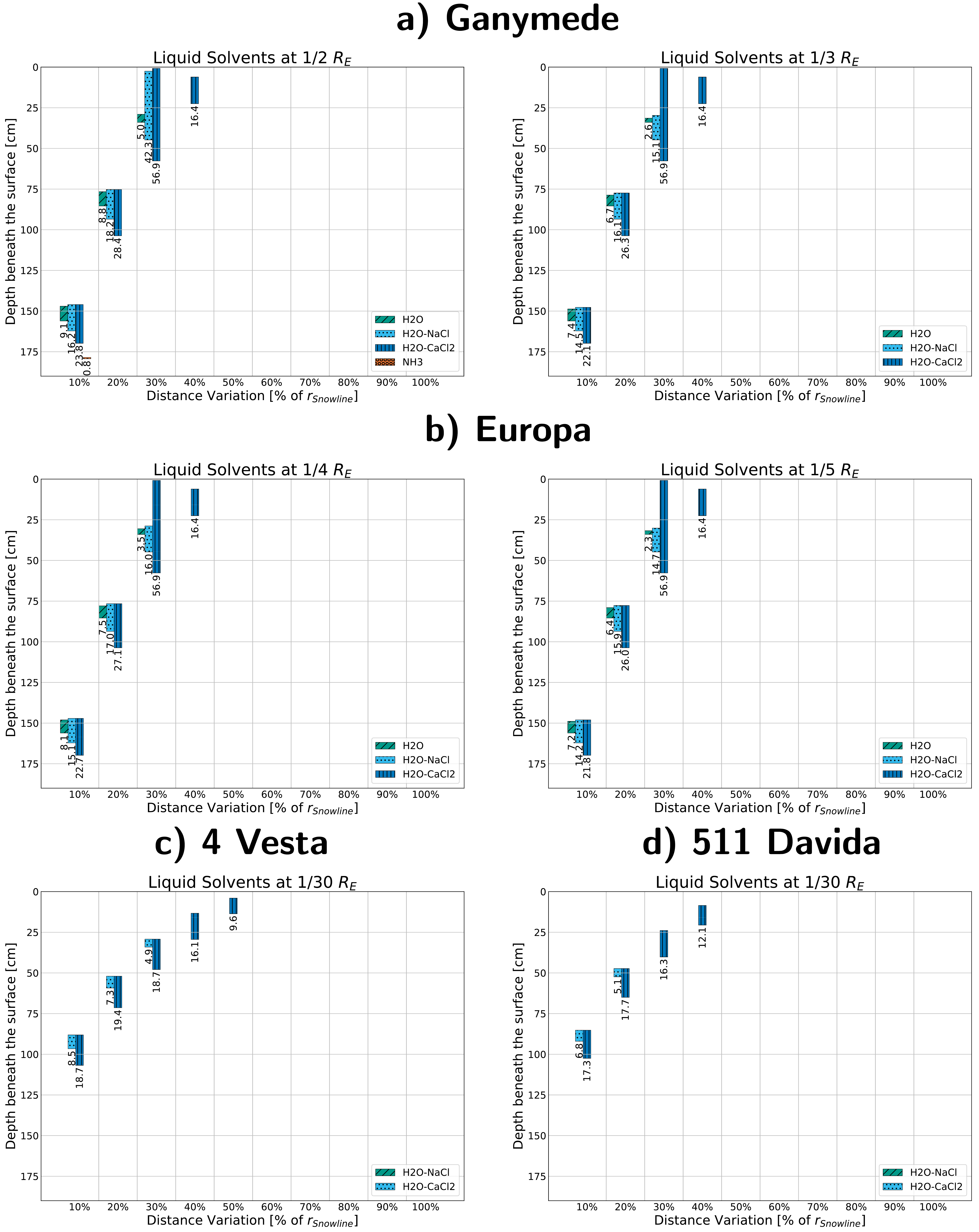}
\par\end{centering}
\caption{\label{fig:rhobodies}Comparisons between different approximations
of smaller bodies with sizes and densities close to their real-life
counterparts, with crust densities $\rho_{surface}$ as seen in Table
\ref{tab:SurfDensity}. a) Jupiter's moon Ganymede ($\rho_{body}=1.936\frac{g}{cm{{}^3}}$)
shown for sizes of $\frac{1}{2}\&\frac{1}{3}R_{earth}$, calculated
using an icy crust. b) Jupiter's moon Europa ($\rho_{body}=3.014\frac{g}{cm{{}^3}}$)
shown for sizes of $\frac{1}{4}\&\frac{1}{5}R_{earth}$, calculated
using an icy crust. c) Asteroid 4 Vesta ($\rho_{body}=3.58\frac{g}{cm{{}^3}}$)
shown for a size of $\frac{1}{30}R_{earth}$, calculated using a rocky,
freshwater crust. d) Asteroid 511 Davida ($\rho_{body}=2.48\frac{g}{cm{{}^3}}$)
shown for a size of $\frac{1}{30}R_{earth}$, calculated using a rocky,
freshwater crust with modified density $\rho_{surface}=\rho_{body}$.}
\end{figure*}

We then did rough approximations of four small real bodies in the
solar system to show the results of a more constraint but realistic
approach to density and size. Simulations were ran after adjusting
$\rho_{body}$(as mentioned in \ref{subsec:Derivation-of-triple})
to the density of Ganymede ($1.936\frac{g}{cm{{}^3}}$), 511 Davida
($2.48\frac{g}{cm{{}^3}}$), Europa ($3.014\frac{g}{cm{{}^3}}$) and
4 Vesta ($3.58\frac{g}{cm{{}^3}}$) (\citealt{Showman1999}, \citealt{CARRY201298})
respectively. We further modified crust density to be equal the body
density for the asteroid 511 Davida, as it is non-differentiated.
Results are shown in Fig. \ref{fig:rhobodies}.

\subsection{Determining the minimum valid model body size\label{subsec:Determining-the-minimum}}

\begin{figure*}
\centering{}\includegraphics[width=1\textwidth]{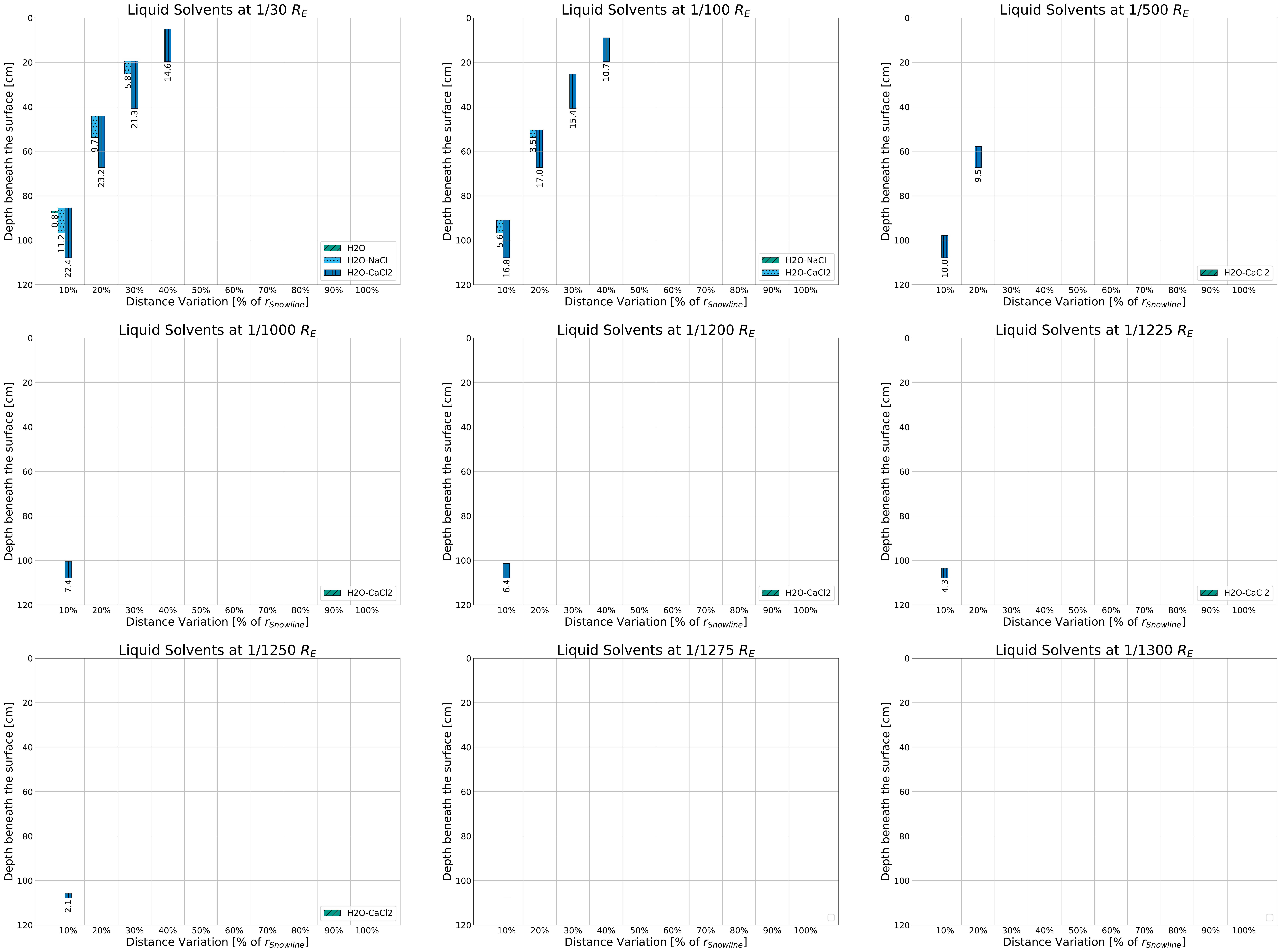}\caption{\label{fig:minR_rocky}Liquid layers on very small bodies using the
standard rocky, freshwater model ($\rho_{surface}=2.2765\frac{g}{cm{{}^3}},\,\rho_{body}=5.51\frac{g}{cm{{}^3}}$).
On distances very close to the source, both forms saltwater can be
liquid on bodies as small as $\frac{1}{100}R_{earth}$, with $CaCl_{2}$-saltwater
even being liquid on bodies as small as $\frac{1}{1250}R_{earth}$.}
\end{figure*}

As a last point of this appendix, we attempt to determine a rough
measure of how small a body within a certain subset of parameters
must be in order to support any form of liquid layers. This has been
done by manually adjusting the model body sizes within the final calculations
of where liquid layers are present using the thermal profile of the
rocky crust model as an input - meaning that these different sizes
were not taken into account during the simulation of said thermal
profiles, again to stay within a certain realistic scope. The results
can be seen in Fig. \ref{fig:minR_rocky}. One can see that, in accordance
with our previous assessments, saltwater is able to persists with
only marginal limitations from a body's size (and therefore the pressure
environment beneath the surface). This results in $CaCl_{2}$-saltwater's
ability to, in our simulations, stay liquid even on bodies as small
as$\frac{1}{1250}R_{E}\approx5km$\footnote{We consider the visible sub-millimeter remnant at$\frac{1}{1275}R_{E}$practically
unviable and a potential artifact of the simulation, due to it being
smaller than the respective stepsize.}. It should be noted that at this size, many other factors may support
or hamper the persistence of liquid solvents that we did not take
into account here, but this serves as a fitting proof-of-concept that,
in environments such as the ones considered here, even bodies as small
as this are worth investigating more closely.

\end{appendix}
\end{document}